# Impact of Fuel Injection Temperature Dynamics on the Stability of Liquid Oxygen-Methane Supercritical Combustion

Abhishek Sharma[1,2], Ashoke De[2,3*], S.Sunil Kumar[1]

[1]*Liquid Propulsion Systems Center, ISRO, Valiamala, 695547, Thiruvananthapuram, India*

[2]*Department of Aerospace Engineering, Indian Institute of Technology Kanpur, 208016, Kanpur, India*

[3]*Department of Sustainable Energy Engineering, Indian Institute of Technology Kanpur, 208016, Kanpur, India*

A crucial factor in the stability of high-pressure rocket-scale combustors is the temperature at which fuel is injected. This study investigates its effect on the stability of supercritical liquid oxygen (LOx)-methane combustion and highlights the impact of shear layer dynamics in cases with lower injection temperatures. The stability features of a rocket-scale combustor operating with multiple injector elements are investigated using a high-fidelity large eddy simulation (LES) framework. The numerical framework combines a flamelet-generated manifold (FGM) combustion model with complex real gas thermodynamics in a scale-resolving simulation setup. It reproduces the non-equilibrium transcritical injection and supercritical combustion characteristics of supercritical methane-oxygen flames. To ascertain the effect of injection temperature on flame and combustor stability, we perform several LES simulations at various methane injection temperatures and produce a stability map. Our analysis shows extremely unstable flame characteristics at lower fuel injection temperatures that are not seen under typical fuel injection circumstances. Below a specific methane injection temperature, LES captures a high-amplitude, self-sustaining instability. It is determined that the combustor becomes unstable below a specific stability boundary temperature. Detailed spectral and dynamic mode decomposition (DMD) analysis of the stable and unstable cases reveals the onset of longitudinal acoustic waves in the combustor. Our thorough investigation pinpoints the instability mechanism, emphasizing that the leading causes of this self-sustaining instability in the combustor are a reduced velocity ratio, fuel buildup, and fuel cut-off occurrences.

## I. INTRODUCTION

With the growing interest in interplanetary missions and reusable launch vehicles, the LOx-methane propellant combination has emerged as the preferred choice for these applications. In recent years, the demand for high-performance and reliable rocket propulsion systems has driven significant interest in LOx-methane engines owing to their advantageous properties, such as high specific impulse and reusability. To achieve both powerful thrust and optimal combustion efficiency, liquid rocket engines operate at extremely high chamber pressures. This requires propellants to be delivered at even higher injection pressures, often in a supercritical or transcritical state. Ensuring combustion performance and stability remains a critical challenge, particularly in

Author to whom correspondence should be addressed. Electronic mail: ashoke@iitk.ac.in

multi-element configurations where interactions between flame dynamics and acoustic modes are pronounced. It is imperative to study LOx-methane combustion to advance the design of reliable rocket engines, minimizing the risk of instability-induced failures and optimizing the performance of next-generation launch vehicles. Aspects of LOx-methane combustion performance and dynamics must be studied to design next-generation launch vehicles. Combustion instability has historically presented a significant obstacle to the advancement of propulsion systems. A coupling of unsteady heat release and the combustion chamber's acoustic eigenmodes causes combustion instability. Fluctuations amplify when heat release is in phase with the acoustic pressure, as first explained by Lord Rayleigh[1]. The instabilities can lead to higher stresses and heat fluxes on the chamber wall and injectors, even leading to the destruction of the engine. It has led to long development times and even the stoppage of development for instability-prone designs. It is known that lower fuel injection temperatures can trigger combustion instability, and there may be a threshold below which the combustor becomes unstable. Any new combustor design must identify this limit, which can only be done by computational analysis or experimental testing. The effect of fuel injection temperature on LOx-hydrogen ($H_2$) combustion stability has been extensively studied experimentally. The characteristics of combustion instability in the LOx-$H_2$ experimental combustor at low hydrogen injection temperatures were thoroughly gathered by Hulka and Hutt[2]. An 89 kN thrust engine with concentric-orifice element injectors was used for the experiments. A variety of geometrical and operational characteristics were tested. There was acoustic instability throughout the hydrogen ramping phase. It was found that the LOx-$H_2$ combustor became unstable due to changes in the hydrogen injection temperature. Tests indicated that the variation of the hydrogen transition temperature is nearly a linear function of the hydrogen-to-oxygen velocity ratio. The hydrogen-to-oxygen injection velocity ratio was identified as a critical stability parameter. A higher injection velocity ratio was one of the most effective methods to stabilize liquid oxygen/hydrogen combustion for concentric-orifice element injectors. The vaporization rate's response to pressure waves was thought to be less pronounced at greater relative velocities between hydrogen and liquid oxygen than at lower relative velocities. Nunome et al.[3] observed similar instability during hydrogen injection temperature ramping tests for a single coaxial injector element combustor. These tests found a stability boundary at 60 K when the hydrogen injection temperature gradually lowered to 50 K. It was found that lowering the temperature at which hydrogen was injected changed the injection velocities, velocity ratio, mixture ratio, and momentum flux ratios, all altering combustion behavior. Furthermore, it emphasizes the significance of the relative velocity ratio as a reduction in the LOx post diameter and, eventually, a lower velocity ratio caused the stability boundary to shift to a lower temperature.

Our literature survey highlights that low hydrogen injection temperatures have been the focus of extensive research on high-frequency combustion instabilities in LOx- $H_2$ engines[4,5]. However, very few studies have specifically investigated the factors affecting the stability of LOx-methane flames, particularly the role of fuel injection temperature. Jensen et al.[6] tested a multi-element LOx-methane experimental combustor at NASA Lewis Research Center. The initial goal was to evaluate the performance of LOx-methane propellants, followed by temperature ramping and bomb tests. The findings showed self-excited instability at the combustor's first transverse (1T) mode. Furthermore, a higher mode was triggered at 14 kHz in the temperature ramping tests, indicating unstable dynamics at lower methane injection temperatures. It was proposed that coupling between the LOx post and the chamber acoustics was the cause of the observed frequency in the temperature ramping tests. Cup burning and notable variations in fuel mass flow rates were thought to be the causes of the 1T instability (5.2 kHz) and the 14 kHz instability. Recently, Martin et al.[7] investigated combustion instabilities in a LOx-LNG combustor and observed similar instability behavior to earlier LOx-hydrogen tests conducted in the same combustor. They identified the flame flashback phenomenon and the injector-chamber coupled response as the leading causes of acoustic instabilities. Kawashima et al.[8] investigated combustion instabilities in LOx-methane subscale combustor tests, linking the instabilities to LOx post geometry and propellant flow conditions. They postulated that injector design significantly impacted stability and that ring vortex generation was induced by Kelvin-Helmholtz instability. It was found that, especially with thinner posts, both injection-coupled instability and Kelvin-Helmholtz instability contribute to creating vortices at the LOx post. Zhang et al.[9] conducted numerous tests on a 600 kN-class LOx-$LCH_4$ engine designed by the Beijing Aerospace Propulsion Institute to evaluate its cooling, performance, and combustion stability. The effect of methane injection temperature on combustion stability was also evaluated. It was determined that the stability limit for methane injection lies around 220 K, below which the engine becomes unstable.

Previous attempts to investigate combustion instability in rocket engines have relied on analytical theories or sub-scale experiments[10–13]. Investigations of combustion instabilities in sub-scale rocket combustors have been carried out experimentally[14–16]. However, it is still challenging to visualize combustion flow fields at high-pressure conditions. Even with the most recent methods, obtaining temperature and species distributions is still challenging. Large pressure fluctuations in combustion instability experiments also pose significant safety risks. Analytical or acoustic equation approaches mainly rely on experiments, simulations, or theoretical models for flame information. On the other hand, a compressible reactive large eddy simulation (LES) algorithm captures instantaneous flow characteristics, vortices, eddies, flame, and acoustic waves. This enables the direct observation of nonlinear limit cycle oscillations and flame response. Recently, combustion instabilities in sub-scale rocket experiments have been effectively studied using a high-fidelity LES framework [17–20]. Experimental investigation of combustion instability at rocket engine-relevant conditions has been pursued at Purdue University over the last decade to generate data and develop LES codes that can predict thermo-acoustic instabilities. However, computational studies on the impact of reduced fuel injection temperatures on stability are limited and, till now, have been conducted on laboratory-scale LOx-$H_2$ combustors only. Literature reports few attempts in this direction. Schmitt et al.[21] simulated DLR combustor tests at 90 K and 45 K hydrogen inlet temperatures, respectively. The study showed dynamic activity in the combustor compared to stable operation observed in the tests at an injection temperature of 45 K. This difference in behavior was attributed to the dynamic fluid conditions that prevailed during the test, as opposed to the static inputs used in simulations. Schmitt et al.[22] also studied JAXA shear coaxial single-element long combustor using an infinitely fast chemistry model at varied hydrogen injection temperatures. Dominant chamber frequency was captured in case of lower hydrogen injection temperature, even though the amplitude was lower than the test data. Otta et al.[23] recently conducted a two-dimensional computational analysis to examine how a lower hydrogen temperature affects a single-element $H_2$-$O_2$ combustor's stability. At a low fuel injection temperature of 80 K, it exhibits severe pressure oscillations and stable combustion for fuel injection at 300 K. It shows the onset of the acoustic wave corresponding to the first longitudinal mode of the combustor. It was discussed that unburned gas tends to accumulate in the far downstream region with low-temperature injection due to delayed combustion. The subsequent heat release from the unburned gas then leads to the development of pressure waves. The effect of a lower oxidizer temperature on the stability of a continuously variable resonant combustor (CVRC) was investigated by Harvazinski and Shimizu[24]. It shows the lower oxidizer temperature increases the amplitude of the pressure oscillations, which was attributed to an increase in the combustion recovery time after the fuel cut-off event.

The primary motivation of this work is to address a significant gap in the field by investigating the combustion dynamics of LOx-methane flames in a rocket-scale combustor with multiple injectors. To the best of our knowledge, no numerical study in the literature has examined the impact of fuel injection temperature on the stability of LOx-methane flames. There is a clear gap in the investigation of how fuel injection temperature affects the stability limits of methane-oxygen supercritical combustion. Specifically, there is a lack of research on the occurrence of combustion instability in multi-injector LOx-methane combustors operating under rocket-relevant conditions. The primary goal of this study is to generate a stability map of LOx-methane supercritical combustion using high-fidelity reactive flow simulations. This stability mapping aims to understand and predict the conditions under which the combustion process remains stable or transitions to instability, which is crucial for designing reliable rocket engines. In this work, fuel injection temperature is considered a crucial parameter, and its impact on the stability of LOx-methane supercritical combustion is explored. This study aims to analyze how variations in fuel injection temperature affect combustion stability by accurately modeling the turbulent and unsteady flow dynamics in a typical multi-injector LOx-methane combustor. Moreover, we thoroughly discuss the potential mechanisms of instability observed under off-nominal injection temperature conditions.

The paper is divided into four main sections. We start with a brief overview of the theoretical background and the numerical framework used to simulate LOx-methane flames, followed by a brief description of the high-fidelity large eddy simulation (LES) framework developed in our earlier work[20], which incorporates real gas thermodynamics and the flamelet-generated manifold (FGM) combustion model. The subsequent section describes the computational domain of a rocket-scale combustor with multiple injector elements. The domain is used to simulate the effects of multi-element flame interactions and flame-wall interactions. Model validation efforts are presented, effectively capturing methane-oxygen flames' injection and combustion dynamics. The validated methodology is then applied to perform several LES simulations on the multi-element domain by varying the methane injection temperature. A detailed overview of the dynamics of the multi-element combustor for different fuel injection temperatures is presented next. The stability characteristics at varied fuel injection temperatures are illustrated using detailed spectral and dynamic mode decomposition analyses. A detailed comparison between unstable and stable cases highlights the key factors influencing the stability of LOx-methane flames.

The concluding section provides a detailed analysis to reveal the instability mechanism at play, highlighting the role of lower velocity ratios, fuel accumulation, and fuel cut-off events as key contributors to the self-sustained instability in the combustor. A critical stability boundary temperature is identified, below which the combustor becomes unstable. A stability map is generated based on fuel injection temperature to offer clear design input for stable LOx-methane engine operation.

## II. THEORETICAL DETAILS AND NUMERICAL FRAMEWORK

### A. Thermodynamic Modelling

Selecting thermodynamic and transport property models is crucial for accurately simulating transcritical injection and supercritical combustion. Oefelein and Yang[25]'s initial study provides an in-depth analysis of high-pressure mixing and combustion processes in liquid rocket engines, highlighting key challenges in transcritical and supercritical mixing. It emphasizes that diffusion-dominated mixing is more critical than atomization in the pre-combustion phase. Our detailed review of phase separation[26] shows that a dense gas/single-phase model is sufficient for simulating LOx-methane injection, mixing, and supercritical combustion. Our review highlights that localized phase separation may occur near the oxygen stream at low temperatures. Still, the more significant part of the LOx-methane flame remains unaffected by phase instability. Numerical studies reveal no significant impact of phase separation, with the dense-gas model yielding identical results simulating benchmark tests. Similarly, remarkable experimental studies by Candel and Singla et al.[27,28] at ONERA showed no visible droplet formation or phase separation during LOx-methane(G2) tests at the Mascotte facility. A recent remarkable experimental study conducted by Boulal et al.[29] at ONERA illustrates the flame structure across sub-critical to supercritical operating pressure ranges. The study reveals the notable absence of a sharp interface resembling droplets, even at a reduced pressure of 0.9 for oxygen injection. We found that phase separation does not affect the high-temperature region of the flame and can be ignored in computational modeling. However, transcritical and supercritical regimes exhibit non-linear behavior in thermodynamic and transport properties, especially near pseudo-critical temperatures. This necessitates using real fluid equations of state (EoS) and departure functions to model thermodynamic properties accurately. Our earlier work[26] provided a detailed evaluation of thermodynamic and transport properties, using the Soave-Redlich-Kwong (SRK) real gas model to accurately compute thermodynamic states for both methane and oxygen across the full range of temperatures and pressures in this study. In our recent work[30], we also assessed various real fluid EoS and Chung's transport model for predicting thermophysical properties in high-pressure LOx-methane engines, confirming the SRK EoS as effective in modeling oxygen and methane thermodynamic states.

In this study, we adopt the same thermodynamic approach to simulate temperature dynamics and the stability of LOx-methane flames. A pseudo-critical method [31] with Van der Waals mixing rules describes mixture properties. The details on the departure function formalism for thermophysical properties can be referred to in our earlier works[20,26]. The importance of accurately describing transport properties under transcritical injection conditions was emphasized in earlier studies. We use the established Chapman–Enskog theory with dense fluid corrections. Chung et al.[32] method is applied to calculate molecular viscosity and thermal conductivity, accounting for gas-like and liquid-like behavior through the viscosity coefficient at pressures and temperatures above and below the critical point.

## B. Numerical Framework

This section provides an overview of the numerical framework used in this study. Combustion instability is an inherently unsteady phenomenon that requires time-resolved, high-fidelity simulations. Large eddy simulation (LES) effectively captures turbulent mixing, unsteady heat release, acoustic wave propagation, and flame-acoustic interactions leading to combustion instability. The coupling between heat release and acoustics relies on resolving turbulent flow, with LES being essential for capturing the large-scale eddies that dominate reacting flow dynamics. In this study, best practices evolved based on our prior works[20,33–35] are employed. A LES framework with real-gas thermodynamics and a high-fidelity combustion model simulates instabilities in a multi-element rocket combustor. This framework effectively resolves turbulence and combustion scales to assess the impact of fuel injection temperature dynamics on LOx-methane flame stability. Our earlier study provides a detailed combustion dynamics framework; however, key governing equations are summarized here for completeness. The large-scale eddies are captured with appropriate mesh resolution, while sub-grid scales are modeled using the dynamic Smagorinsky-Lilly model by Germano et al.[36]. We present the compressible Favre-filtered governing equations used in the numerical framework, with further details available in our previous works[20,33].

**Filtered continuity:**

$$\frac{\partial \bar{\rho}}{\partial t} + \frac{\partial}{\partial x_i}(\bar{\rho}\tilde{u}_i) = 0 \qquad (1)$$

Where $\rho$ is density and $u$ is the velocity component.

**Filtered momentum:**

$$\frac{\partial}{\partial t}(\bar{\rho}\tilde{u}_i) + \frac{\partial}{\partial x_j}(\bar{\rho}\tilde{u}_i\tilde{u}_j) = \frac{\partial}{\partial x_j}(\widetilde{\sigma_{ij}}) - \frac{\partial \bar{p}}{\partial x_i} - \frac{\partial \tau_{ij}}{\partial x_j} \qquad (2)$$

Where $p$ is pressure, $\sigma_{ij}$ is the stress tensor due to molecular viscosity defined by:

$$\sigma_{ij} \equiv \left[\mu\left(\frac{\partial\, \overline{u_i}}{\partial x_j} + \frac{\partial\, \overline{u_j}}{\partial x_i}\right)\right] - \frac{2}{3}\mu\frac{\partial \overline{u_l}}{\partial x_l}\delta_{ij} \qquad (3)$$

$\mu$ is molecular viscosity and $\delta_{ij}$ is Kronecker delta.

The compressible form of the sub-grid stress tensor $\tau_{ij}$ is defined as:

$$\tau_{ij} \equiv \rho\widetilde{u_i u_j} - \rho\, \widetilde{u_i}\widetilde{u_j} \qquad (4)$$

**Energy/Enthalpy equation:**

$$\frac{\partial\rho\widetilde{h_s}}{\partial t} + \frac{\partial\rho\widetilde{u_i}\widetilde{h_s}}{\partial x_i} - \frac{\partial\bar{p}}{\partial t} - \widetilde{u_j}\frac{\partial\bar{p}}{\partial x_i} - \frac{\partial}{\partial x_i}\left(\lambda\frac{\partial\tilde{T}}{\partial x_i}\right) \tag{5}$$

$$= -\frac{\partial}{\partial x_j}\left[\bar{\rho}\left(\widetilde{u_i h_s} - \widetilde{u_i}\widetilde{h_s}\right)\right]$$

Where $h_s$& $\lambda$ are the sensible enthalpy and thermal conductivity, respectively. The subgrid enthalpy flux term in the filtered energy equation is approximated using the gradient hypothesis:

$$\bar{\rho}\left(\widetilde{u_i h_s} - \widetilde{u_i}\widetilde{h_s}\right) = -\frac{\mu_{SGS}\, c_p}{Pr_{SGS}}\frac{\partial\tilde{T}}{\partial x_j} \tag{6}$$

$\mu_{SGS}$ is the subgrid viscosity and $Pr_{SGS}$ is the subgrid Prandtl number and is equal to 0.85. $C_p$ is specific heat capacity, and $T$ is temperature. The subgrid-scale stresses resulting from the filtering operation are modeled using sub-grid scale models, which employ the Bousinessq hypothesis used in RANS models.

$$\tau_{ij} = -2\mu_t\bar{S}_{ij} \tag{7}$$

$$\bar{S}_{ij} \equiv \frac{1}{2}\left(\frac{\partial\,\bar{u}_i}{\partial x_j} + \frac{\partial\bar{u}_j}{\partial x_i}\right) \tag{8}$$

Where $\mu_t$ is the subgrid-scale turbulent viscosity. $\bar{S}_{ij}$ is the rate-of-strain tensor for the resolved scale

A modified configuration of sub-grid stress tensor $\tau_{ij}$ for compressible flows is used:

$$\tau_{ij} = \tau_{ij} - \frac{1}{3}\tau_{kk}\delta_{ij} + \frac{1}{3}\tau_{kk}\delta_{ij} \tag{9}$$

$(\tau_{ij} - \frac{1}{3}\tau_{kk}\delta_{ij})$ is a deviatoric part of subgrid-scale stress and $\frac{1}{3}\tau_{kk}\delta_{ij}$ is an isotropic part. The deviatoric part of the subgrid-scale stress tensor is modeled using a compressible form of the Smagorinsky model:

$$\tau_{ij} - \frac{1}{3}\tau_{kk}\delta_{ij} = 2\mu_t\,(\bar{S}_{ij} - \frac{1}{3}S_{kk}\delta_{ij}) \tag{10}$$

The eddy viscosity is modeled by the Smagorinsky-Lilly model given as:

$$\mu_t = \rho\, L_s^2|\bar{S}| \tag{11}$$

Where $L_s$ is the mixing length for the subgrid scales and

$$|\bar{S}| = \sqrt{2\bar{S}_{ij}\bar{S}_{ij}} \tag{12}$$

$L_s$is computed using, $L_s = \min\,(kd\,, C_s\Delta)$ where $C_s$ is Smagorinsky constant and $\Delta$ is the local grid scale. The dynamic version of the Smagorinsky-Lilly model by Germano et al.[36] is used in this study in which $C_s$ is dynamically computed based on the information provided by the resolved scales of motion.

## C. Combustion Model

To accurately model the effect of fuel injection temperature on the stability of supercritical LOx-methane flames, the transient nature of the flame must be captured appropriately. Under supercritical conditions, the smallest turbulent eddies can penetrate the flame, leading to local extinction and re-ignition. Our earlier work [26] provides an assessment of different combustion models for LOx-$CH_4$ supercritical combustion, highlighting the efficacy of the flamelet-generated manifold (FGM) approach for modeling high-pressure LOx-CH4 flames. The FGM[37,38] combustion model incorporates conserved scalars such as the mixture fraction (f) and progress variable (c), effectively reducing the number of equations to solve by transforming the problem from physical to mixture fraction space. Recent studies[39–42] have successfully utilized the FGM approach, solving diffusion flamelet equations and tabulating scalars such as temperature and species mass fraction by integrating their statistical distribution using a probability density function (PDF). Our earlier assessment[26] provided a comprehensive investigation of reactive flow models and kinetic mechanisms. We conducted a detailed analysis of five different kinetic mechanisms to evaluate their ability to accurately capture LOx-methane combustion. It clearly demonstrates the efficacy of the FGM combustion model. While we concluded in our previous work that even a compact JL-R mechanism is effective for capturing LOx-methane combustion, based on comparisons of simulated flame structures with experimental observations, we opted to use the DLR-Kong mechanism[43] for all our LES cases in this study. This mechanism, developed by the German Aerospace Center (DLR) for rocket thrust chambers, comprises 23 species and 51 reactions and showed good agreement with the experimental flame shape. The 23-step Chemkin mechanism is used to create non-adiabatic flamelets for pre-processing mean quantities in the PDF table. This study generates a two-dimensional manifold in reaction progress and mixture fraction space by solving diffusion flamelet equation and converting species mass fraction in terms of reaction progress variable. A total number of 158 flamelets are generated using the inbuilt utility. Automatic grid resolution of flamelet table with a maximum number of 100 points in both the mean mixture fraction and mean progress variable space is employed. Compressibility effects are enabled in the FGM to incorporate pressure oscillations/acoustic response on density/solution variables. In this framework, the transport equations for both control variables, i.e. mixture fraction $f$, and the un-normalized progress variable, $Y_c$ is solved along with basic governing equations. The Favre averaged transport equation for the mixture fraction is given by:

$$\frac{\partial}{\partial t}\left(\rho\tilde{f}\right)+\nabla.\left(\rho\vec{v}\tilde{f}\right)=\nabla.\left(\frac{\mu_{eff}}{\sigma_t}\nabla\tilde{f}\right)+S_m \qquad (13)$$

where, $S_m$ is only due to the mass transfer into the gas phase from liquid droplets. $\mu_{\text{eff}}$ is the effective viscosity composed of a laminar ($\mu_l$) and a turbulent contribution ($\mu_t$). In this FGM framework, the transport equation for $Y_c$ is solved, instead of normalized progress variable, $c$ which has additional terms and is not well defined at the oxidizer boundaries. The solution of $Y_c$ brings added advantage to model the flame quenching naturally with its definition.

$$\frac{\partial \rho \widetilde{Y_c}}{\partial t} + \frac{\partial}{\partial x_i}\left(\rho \tilde{u}_i \widetilde{Y_c}\right) = \frac{\partial}{\partial x_i}\left(D_{eff} \frac{\partial \widetilde{Y_c}}{\partial x_i}\right) + \bar{S}_{Y_c} \quad (14)$$

$\bar{S}_{Y_c}$ is modeled as a finite rate source term taken from the flamelet library using c and f as control variables. This source term $\bar{S}_{Y_c}$ determines the turbulent flame position and is calculated as:

$$\bar{S}_{Y_c} = \bar{\rho} \iint S_{FR}(c,f) p(c,f)\, dc\, df = \bar{S}_{FR} \quad (15)$$

The $p(c,f)$ is a joint PDF specified as the product of two beta PDFs. In this study, the variance of the un-normalized reaction progress variable is modeled using the algebraic formulation, assuming the equilibrium of the generation and dissipation of variance at the sub-grid scale.

$$\tilde{Y}_c'^2 = C_{var} \frac{l_{turb}^2}{Sc_t} \left(\nabla \widetilde{Y_c}\right)^2 \quad (16)$$

Where $l_{turb}$ is the turbulent length scale and $C_{var}$ is the constant with a value assigned as 0.1 in this formulation. The turbulent Schmidt number is used as 0.7 in this study. Similarly, the transport equation for mixture fraction variance is not solved; instead, it is modeled by an algebraic equation given as:

$$\widetilde{f'^2} = C_{var} L_s^2 \left|\nabla \bar{f}\right|^2 \quad (17)$$

The constant $C_{var}$ is computed dynamically based on the dynamic version of the Smagorinsky-Lilly SGS model. Where $L_s$ is the mixing length for the subgrid scales and is computed using, $L_s = \min\,(kd\,, C_s\Delta)$ where $C_s$ is Smagorinsky constant and $\Delta$ is the local grid scale. The constant $C_s$ is also computed dynamically based on the information provided by the resolved scales of motion. Turbulence-chemistry interactions are modeled with an assumed shaped probability density function (PDF) approach to compute average values. Density-weighted mean scalars ($\bar{\phi}$), are calculated from the probability density function (PDF) of f and c with the beta distribution. The mean quantities are tabulated in a look-up table against a specified value of 5 parameters, such as $\bar{f}$, $\overline{f'^2}$, mean $\bar{c}$, variance $\overline{c'^2}$ of progress variable respectively and mean enthalpy, $\bar{H}$. In this study, to reduce the computational cost of generating a five-dimensional table, it is assumed that the heat loss or gain by the system only has a negligible effect on species mass fraction, and mean scalars other than species mass fraction can be computed using the average progress variable. A different set of look-up tables for species mass fraction $\bar{\phi}(\bar{f}, \overline{f'^2}, \bar{c}, \overline{c'^2},)$ and remaining scalars affected by enthalpy change $\bar{\phi}(\bar{f}, \overline{f'^2}, \bar{c}, \bar{H})$ is generated respectively.

## III. COMPUTATIONAL DOMAIN

The LES methodology is applied to capture the dynamics of fuel injection temperature in a multi-injector element combustor operating at supercritical pressure. The combustion instability mechanism in such a combustor is complex due to the interactions among the injectors, the combustor flow, and the flames. We utilize a rocket-scale combustor domain with seven injector elements, precisely seven bi-directional swirl coaxial injectors. These injectors are preferred for their ability to handle large flow rates, deliver higher thrust per element, and achieve greater combustion efficiency. This multi-element combustor domain simulates the injection and mixing processes typical of a full-scale rocket engine. This section describes the computational domain and boundary conditions used in this multi-element LES. The combustor features an axial length 13 times the throat diameter and a chamber diameter 4.18 times the throat diameter. The injector plate accommodates seven elements: one injector at the center and six others arranged hexagonally from the center. Figure 1 shows the computational domain featuring seven swirl coaxial injectors truncated at the throat location. A detailed cutaway view of the injector illustrates methane and oxygen's entry and flow paths. The propellants enter the combustor as separate non-premixed streams, undergoing rapid turbulent mixing at the injector exit region. Each injector element is a bi-directional swirl coaxial design, with oxygen flowing through the central path and methane through an annular path surrounding the oxygen. The tangential entry of oxygen and the presence of helical vanes in the methane's axial passage achieve the bi-directional swirl, inducing a swirl motion. A separate cut view of the injector depicts methane and oxygen's entry and flow path. The oxygen path axial length is 11.3 $R_n$, and the coaxial methane flow path length is 12.25 $R_n$, where $R_n$ is the oxygen post radius. A positive recess of 0.6 $R_n$ is provided to the oxygen post to enhance mixing.

In this study, the LOx-methane injection conditions closely approximate actual rocket engine operating conditions. The flow rate per injector element is set to maintain a nominal chamber pressure of approximately 62 bar. Oxygen is injected in a transcritical state with an injection pressure of 70 bar and a temperature of 83K. In contrast, methane injection temperature varies from a nominal value of 233 K to as low as 200 K. The injection densities are 88 kg/m³ for methane and 1190 kg/m³ for oxygen. A mass flow inlet boundary condition is applied at the inlets of the methane and oxygen injectors. The walls are treated as adiabatic with a no-slip condition imposed. The domain is truncated at the throat, automatically providing acoustically fully reflective boundary conditions due to the choked throat condition.

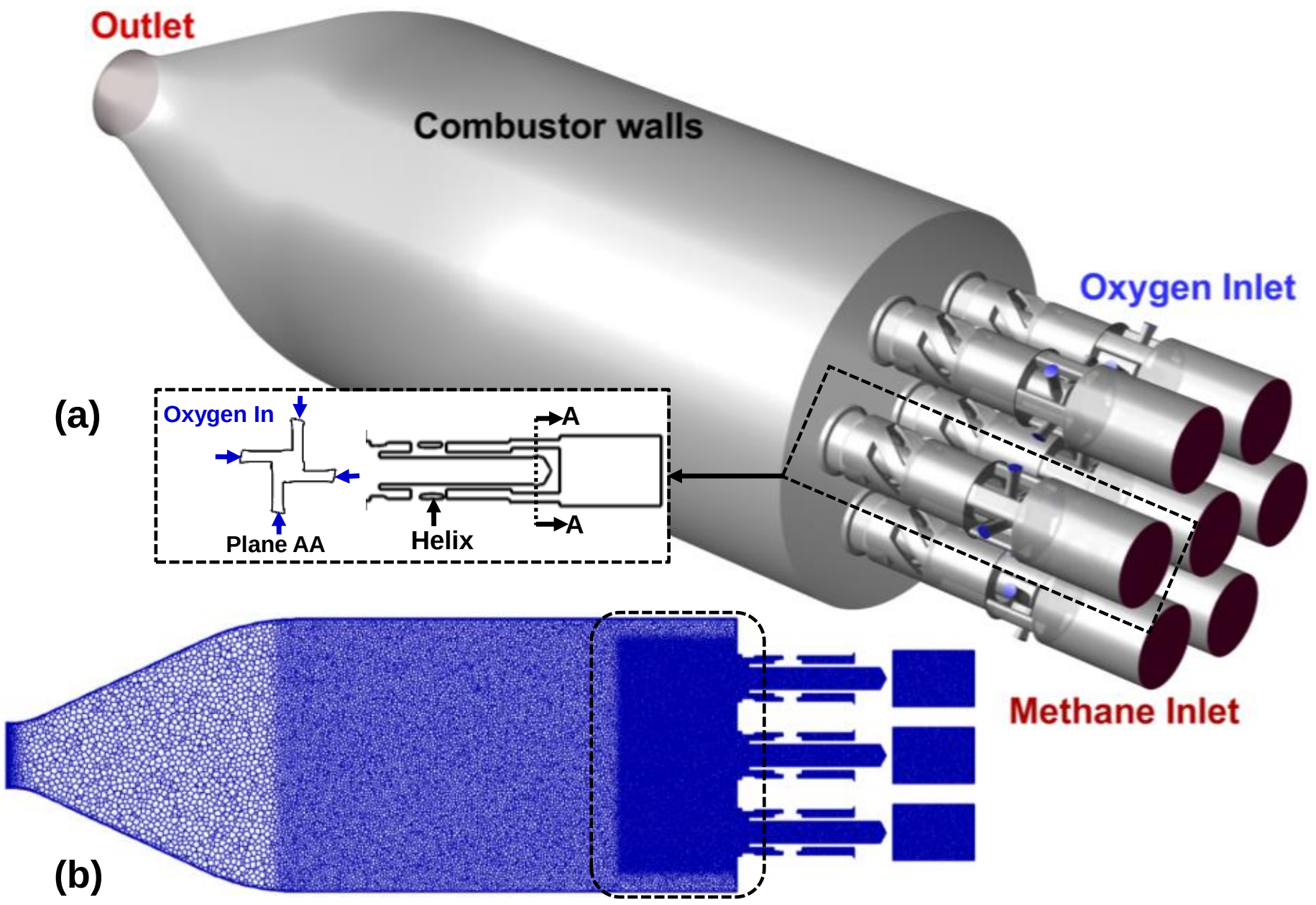

**FIG.1.** Computational domain- Multi-Element Combustor

We employ the mesh methodology developed in our previous work[20]. A suitably refined mesh is created around the injector outlet region to resolve more than 80% of turbulent scales. The mesh quality meets the specifications outlined in the previous work[20], which ensures a high-quality LES mesh near the injector exit that captures steep gradients. This refined mesh features a higher cell density at regions of high gradients, with a total grid size of 6.5 million cells. In this study, a refined grid with 6.5 million cells is used for LES. Figure 1 (b) displays the mesh view at the cut section of the combustor. A higher number of cells are packed close to the injector outlet region, with minimum cell sizes of 0.22 mm, as shown under the marked box. A LES resolution criterion is also used to determine the adequacy of grid size, which can resolve more than 80% of turbulent scales at comparatively lower computational expense. The grid quality is checked by the LES resolution/IQ criterion defined as:

$$LES_{resolution} = \frac{k_{resolved}}{k_{resolved} + k_{SGS}} \quad (18)$$

where $k_{resolved}$ is the resolved kinetic energy expressed as:

$$k_{resolved} = 0.5[(u - \bar{u})^2 + (v - \bar{v})^2 + (w - \bar{w})^2] \quad (19)$$

with $u$, $v$, $w$ the three components of the flow velocity, and the overbar denotes time-averaging.

$k_{SGS}$ expresses the unresolved sub-grid scale (SGS) turbulent kinetic energy, where $\mu_t$ is turbulent viscosity and $c_{var}$ is the dynamic constant.

$$k_{SGS} = 1.5\left(\frac{\mu_t}{\rho c_{var}\Delta_f}\right)^2, \text{constant } \Delta_f \text{ is calculated by } (Cell\ volume)^{1/3} \quad (20)$$

The low value of modeled sub-grid scale viscosity in the core flow/flame region shows the suitability of the grid resolution, resolving a minimum of 80% of turbulent scales in the domain. Figure 2 shows the LES resolution of the utilized mesh, indicating that the ratio of resolved turbulent kinetic energy (TKE) to total TKE exceeds 0.9. This corresponds to an overall LES IQ close to 1 across the domain, particularly in the region downstream of the injector exit, indicating that the mesh used for the LES cases is well-resolved.

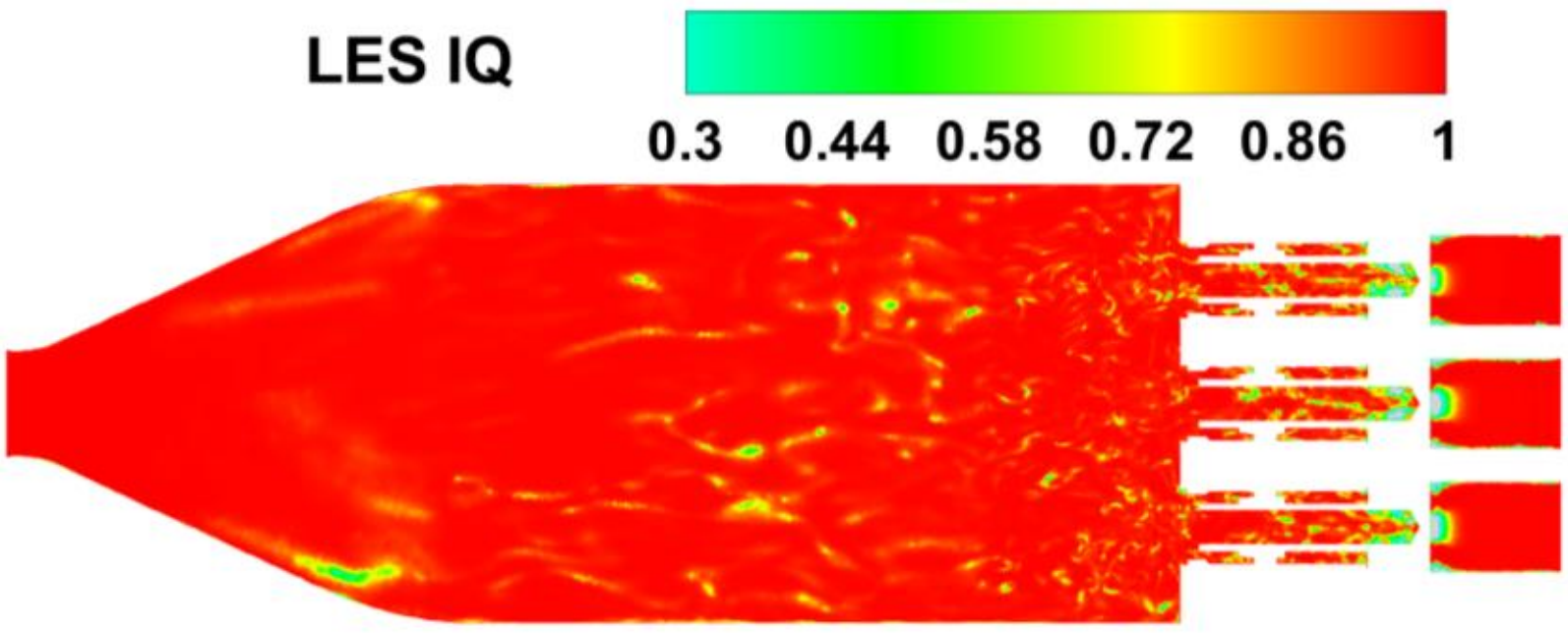

**FIG.2.** LES resolution for temperature case simulations

A time step size sufficient for capturing unsteady dynamics allows for the temporal resolution of injector flow and combustion dynamics without significant numerical dissipation. LES is conducted for a sufficient duration to ensure an initial condition-independent solution and to gather ample unsteady statistics for spectral analysis. The numerical solution techniques developed earlier are utilized in this work. However, the numerical stiffness associated with the combined solution of the real gas and the FGM model within the LES framework and higher-order discretization poses a significant challenge.

All governing equations are implicitly filtered using the finite-volume methodology in ANSYS Fluent[44]. Spatial discretization employs second-order bounded schemes, while time integration utilizes a bounded second-order implicit method. LES is conducted over multiple acoustic cycles to ensure independent results for initial conditions and to gather unsteady pressure statistics for spectral analysis. Further details on the LES setup can be found in our earlier works[20,33]. The LES is performed on 560 direct water-cooled Intel Broadwell cores in a high-performance computing facility.

## IV. MODEL VALIDATION STUDIES

We utilized a validated numerical methodology outlined in our previous work[20,26], based on the Mascotte chamber G2 test case by Singla et al.[27]. This test was simulated to verify the accuracy of the FGM model within a real gas LES framework. The Mascotte test provides an optical view of a LOx-methane flame under supercritical conditions. It is a key benchmark for evaluating numerical models of transcritical injection and supercritical combustion in rocket engines. In the G2 setup, the combustion chamber operates at 5.61 MPa, with oxygen injected in a transcritical state at 85 K and methane in a supercritical state. The chamber features a shear coaxial injector in a rectangular combustor, with oxygen entering the core and methane through the coaxial annulus. In the validation study, the time-averaged OH concentration from the LES closely matches the Abel-transformed OH image from the G2 test, showing strong agreement in axial flame length and radial expansion. Comprehensive details on the validated methodology and computational setup are available in our previous work and included in Annexure A for reference and completeness. We have utilized the Mascotte test case to validate global flame characteristics, while the dynamics part of the code was validated using Purdue University's Continuously Variable Resonance Combustor (CVRC). Our recent work[33] outlines the computational setup and methodology developed to capture self-excited instability. A brief introduction to this validation case is provided here completeness. Yu et al.[14] developed a self-excited single-element rocket combustor to study spontaneous instabilities in a sub-scale rocket chamber. The setup allows for a detailed numerical investigation of the coupling between acoustics and combustion. The instability modes in the CVRC are tuned by adjusting the length of the oxidizer post, which alters the location of the choked inlet in the oxidizer tube. We developed a reactive flow LES modeling framework to capture self-excited and sustained thermoacoustic instability. The model replicated the instantaneous flow field and combustion dynamics observed in CVRC tests, capturing limit-cycle pressure oscillations with dominant frequencies (1L and harmonics), closely matching experimental data. Further details on this study can be found in our work[33,34]. This work uses the Mascotte and CVRC validation schemes to capture the effect of temperature dynamics on the stability of the LOx-methane rocket-scale combustor. Our validation studies also demonstrate that the numerical methodology developed for benchmark cases can be effectively applied to multi-element combustor configurations.

## V. RESULTS AND DISCUSSION

This work emphasizes simulating the effect of variable fuel injection temperatures on combustion stability, revealing the combustor's dynamic response to these variations. It provides insights into flame interaction and dynamics within a multi-injector element setup, examining injector flow physics, the impact of key design parameters, and mechanisms driving pressure fluctuations. The combustor's flow and flame dynamics are initially evaluated through instantaneous LES contours, followed by a detailed spectral and dynamic mode decomposition analysis of probe point and spatial pressure data. The multi-element flame features are first presented for the nominal temperature case, followed by detailed illustrations for other temperature cases. Probable instability mechanisms in unstable cases are discussed in depth, with spectral and dynamic mode decomposition analysis highlighting pressure dynamics in both stable and unstable scenarios. Finally, additional large eddy simulations explore the instability mechanisms further, and a stability map based on different temperature cases is generated.

### A. Injection Temperature Cases- Instantaneous Features

We have simulated seven different fuel injection temperature cases to understand the impact of injection temperature on combustor stability. Computations are performed by varying the fuel injection temperature while keeping the fuel mass flow rate identical in each case. A hot test scenario is simulated through LES in which the overall/global oxidizer to fuel mass ratio (mixture ratio, O/F) is kept constant, but the fuel temperature is ramped up and down. The overall mixture ratio is kept constant in all temperature simulations; however, local changes in the ratio are expected, which could lead to high-amplitude pressure activity and limit cycle oscillations. Fuel injection temperature varies from the nominal temperature of 233 K to the minimum operating temperature of 200 K to generate an overall stability map of this combustor. LES simulations are conducted within these bounds for fuel injection temperatures of 225 K, 221 K, 219 K, 217 K, and 210 K.

We first present the instantaneous flow and flame features for the fuel injection temperature case with stable pressure dynamics. Understanding the multi-element flame structure is necessary before analyzing the impact of fuel injection temperature on combustion stability. Instantaneous contour plots at a simulation time instant of 0.05s reveal significant flow and flame characteristics. Figure 3 displays the instantaneous flame temperature contour at an axial cut plane, illustrating the evolution of supercritical LOx-methane flames, including flame-flame and flame-wall interactions within this multi-injector setup. The temperature contour shows a flame anchored at the end of the oxygen post in the recess region, with high temperatures observed in the shear layer between the oxidizer and fuel near the injector exit, corresponding to stoichiometric conditions. A corrugated flame structure appears in the shear layer, evolving into intense combustion and homogeneous flame mixing further downstream. High-temperature turbulent eddies break away from individual injector flames and merge, forming a well-distributed high-temperature zone. Figure 3(b) shows the radial distribution of instantaneous flame temperature at planes downstream of the injector exit.

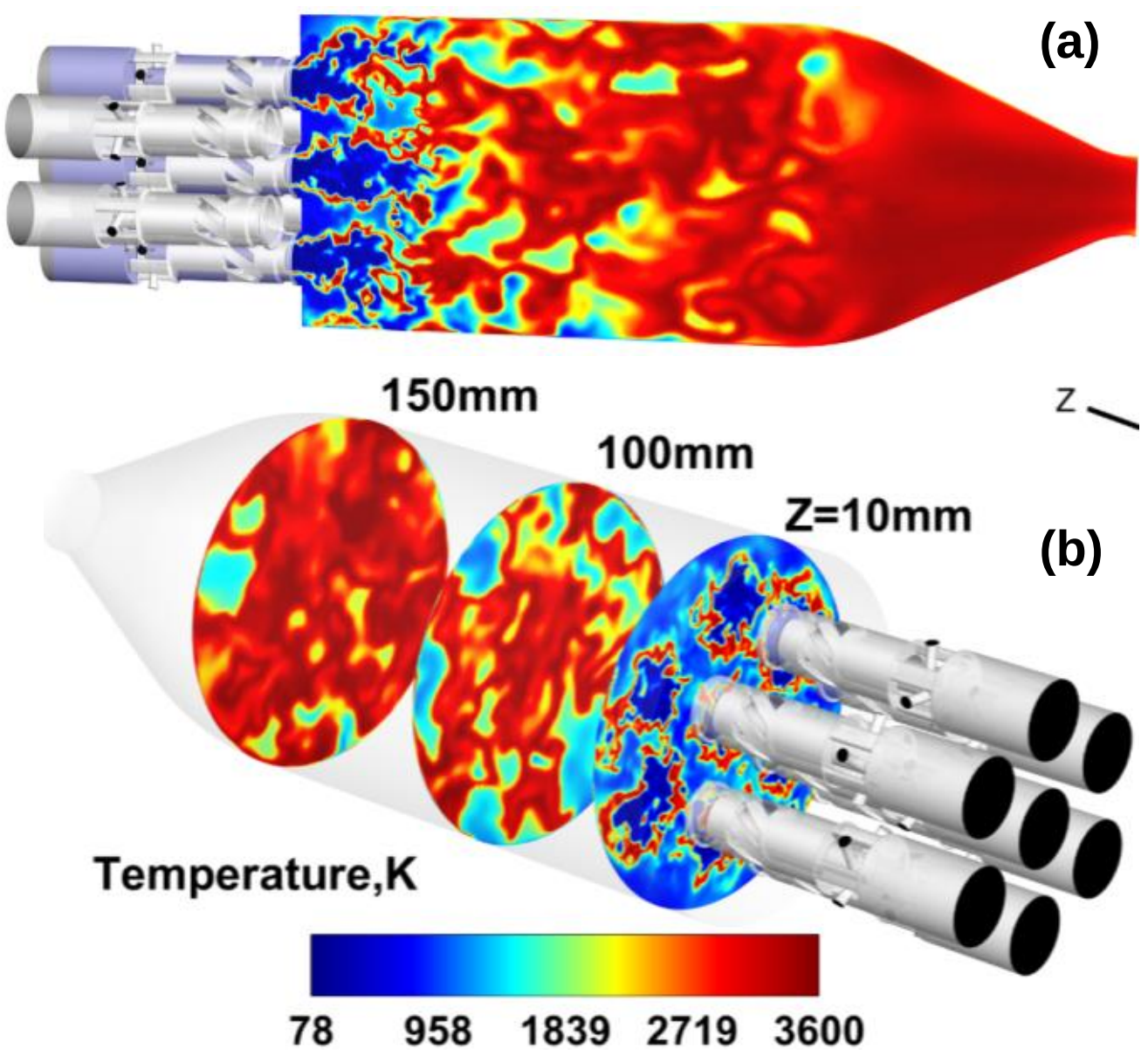

**FIG.3.** Instantaneous temperature variations at (a) axial cut plane, (b) different radial planes

Figure 4(a) depicts the multi-element flame interaction, displaying the mean temperature iso-surface at 1500 K. The flame initiates in the injector recess region and then undergoes radial expansion, followed by flame-flame and flame-wall interactions. The multi-element flame structure can be illustrated through temperature contours, whereas combustion dynamics are assessed through the evolution of fluctuating pressure waves in the combustor. The temporal evolution of pressure can indicate the onset of dynamic activity within the combustor. Figure 4(b) shows the radial distribution of instantaneous absolute pressure at planes downstream of the injector exit. The pressure contour displays the tangential acoustic wave in the combustor, exhibiting a typical pattern with maximum pressure near the combustor wall and lower values at the chamber center. This highlights the formation of tangential mode pressure oscillation at a radial plane, Z=100mm. At the plane, Z=10mm, the radial distribution of pressure shows small zones of high and low pressure distributed throughout the plane. The instantaneous absolute pressure contours illustrate the variation of pressure along the transverse axis of the chamber near the injector exit. This variation depicts the transverse movement of a pressure wave, with high pressure near the bottom wall at one instant and near the top wall at another time. Although the amplitude of transverse acoustic waves is extremely low in this nominal case, simulation highlights the capability to capture the onset of instability at off-nominal operating conditions.

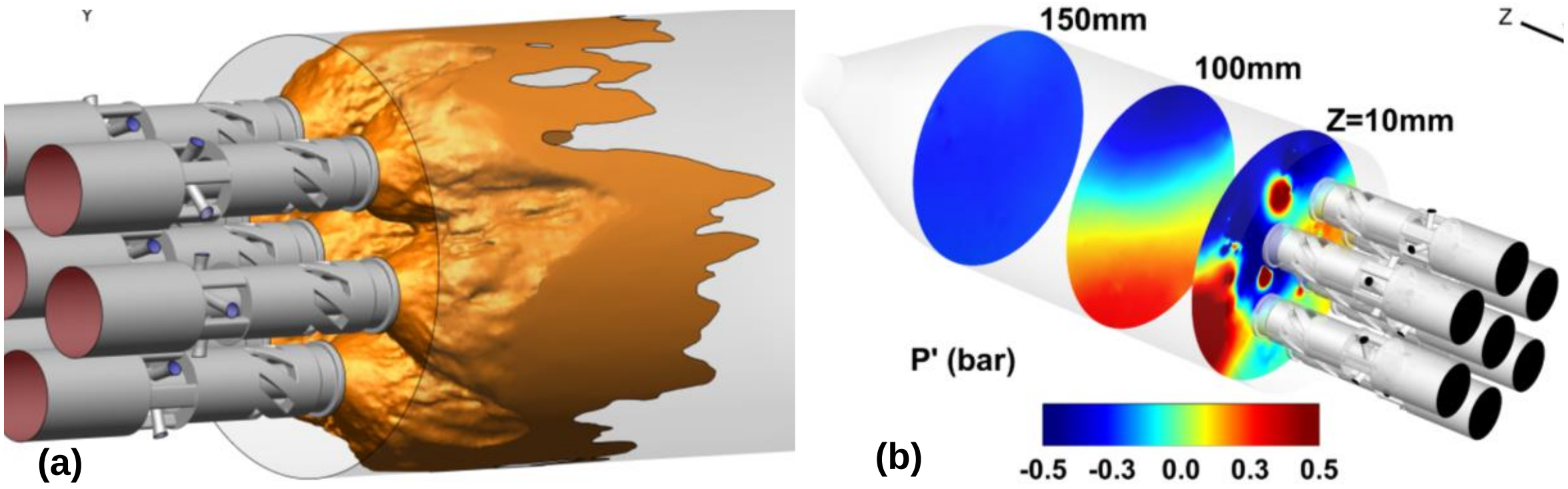

**FIG.4.** (a) Flame temperature iso-surface map, and (b) absolute pressure variation at different radial planes)

A comparative assessment of pressure activity in the combustor at different injection temperatures is discussed first, followed by a detailed analysis of flow and flame parameters for stable and unstable cases. We compare stable and unstable cases through flame characteristics and spectral and dynamic mode decomposition analysis. Unsteady flame features and pressure statistics are analyzed to identify the underlying mechanisms responsible for the onset of high-amplitude pressure fluctuations at off-nominal temperature conditions. A raw pressure trace comparison across different injection temperature cases provides a quick overview of combustor stability. Figure 5 shows the pressure probe locations within the combustor plane. Pressure data is collected at various probe locations throughout the combustor, with nine probes positioned in the combustor region to capture unsteady pressure, six in the oxygen tubes, and two in the methane entry path. The injector probes are labeled as IP, chamber probes as CP, and methane inlet probes as IM. Multiple probes are used to gather unsteady statistics and analyze the dynamic mechanisms in this multi-element configuration. The simulation runs for 40 flow-through times before data collection begins to minimize the effects of initial transients. Absolute pressure is recorded at a sampling frequency of 0.1MHz. Data is collected for over 80ms, corresponding to more than 300 cycles for the first longitudinal (1L) mode and 600 cycles for the first transverse (1T) mode.

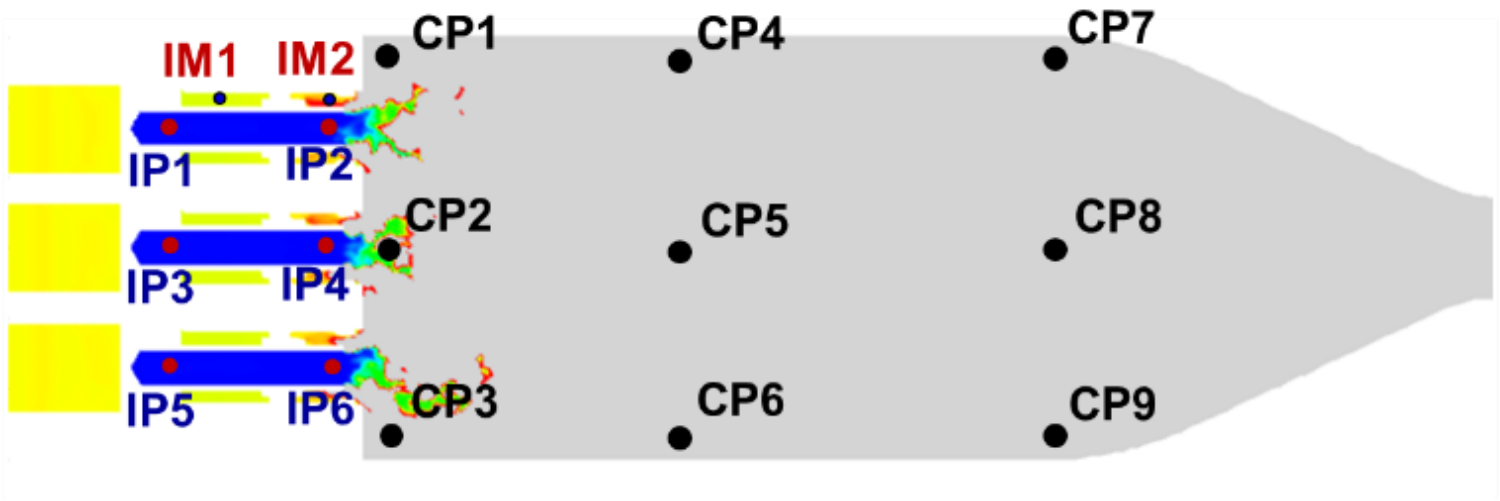

**FIG.5.** Pressure probe location in combustor

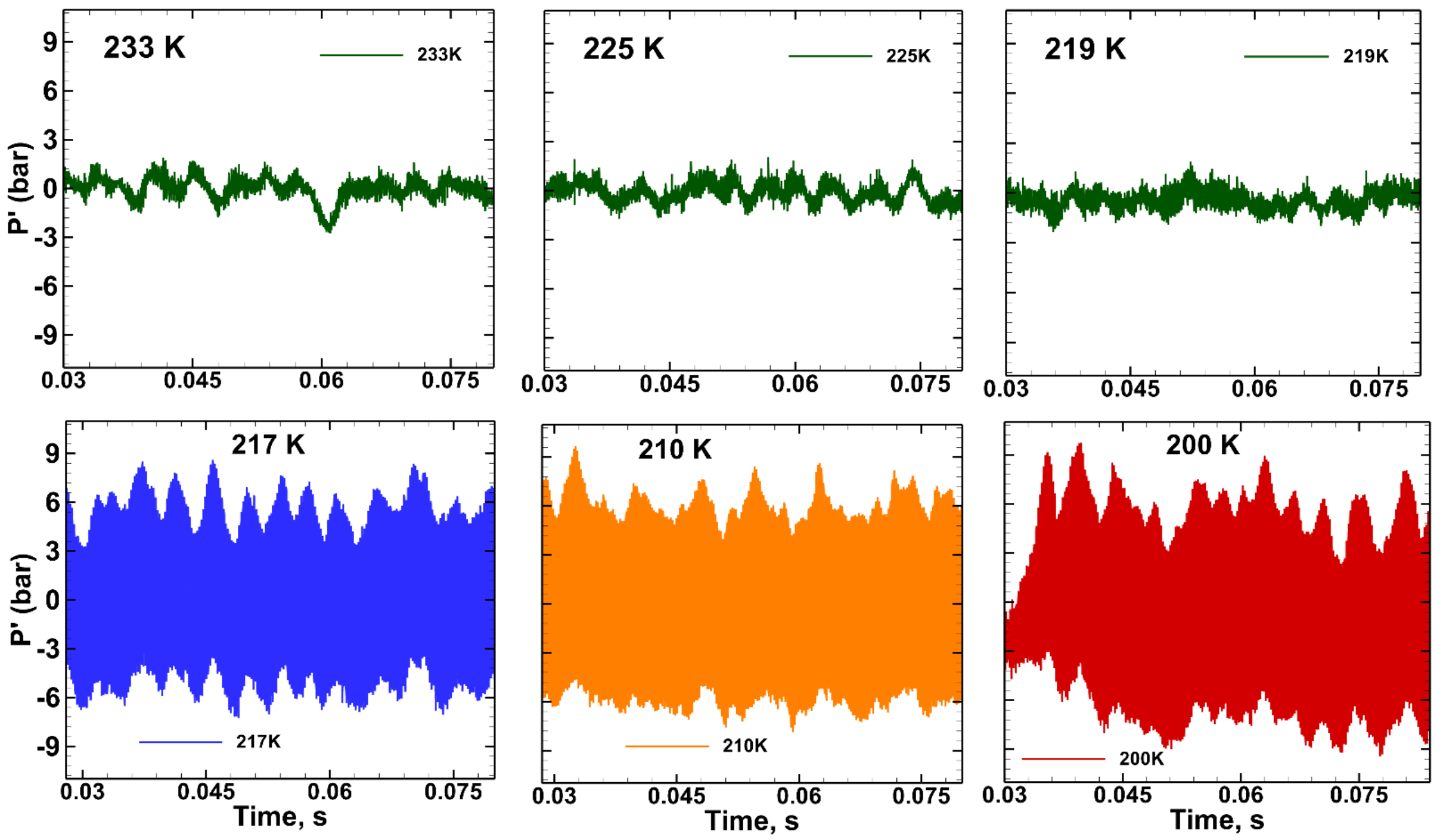

**FIG.6.** Pressure fluctuation for different temperature cases

Figure 6 provides a combined view of pressure fluctuations at the CP1 probe for all temperature cases. A direct comparison of pressure traces at the CP1 probe compares stability features across temperature simulations. It displays low-amplitude fluctuations from an injection temperature of 233 K to 219 K, while high-amplitude pressure activity is observed from 217 K and below. A self-sustained instability appears at temperatures below 219 K. The intense pressure activity reveals limit cycle oscillations that persist until the end of the simulation, displaying the typical characteristics of high-frequency combustion instability. Unstable cases show high peak-to-peak amplitudes, whereas stable cases exhibit very low amplitude fluctuations. The 200 K injection temperature case demonstrates the transition from a stable to an unstable state. The unstable state after 0.031 s exhibits limit cycle oscillations, as shown in the pressure plots for both 217 K and 210 K simulations. Figure 7(a) displays a comparison of pressure fluctuations at the CP1 probe for the maximum (233 K) and minimum (200 K) temperature conditions. A direct pressure trace comparison indicates the onset of instability at 0.31 s for the 200 K case. Figure 7(b) reports a zoomed view of the stable to unstable case comparison in a time window of 1.2ms. It shows a peak-to-peak amplitude near 15 bar for the unstable (200 K) case, while the 233 K case displays stable operation with extremely low-amplitude pressure fluctuations. The unstable case demonstrates self-sustained limit cycle oscillations after transition, while the high-temperature case remains stable. This comparison of pressure oscillations indicates the onset of instability in low-temperature cases, revealing a minimum cut-off injection temperature below which the combustor becomes unstable. A detailed exploration of the mechanism behind the instability observed at lower fuel injection temperatures is thoroughly discussed in the following sections.

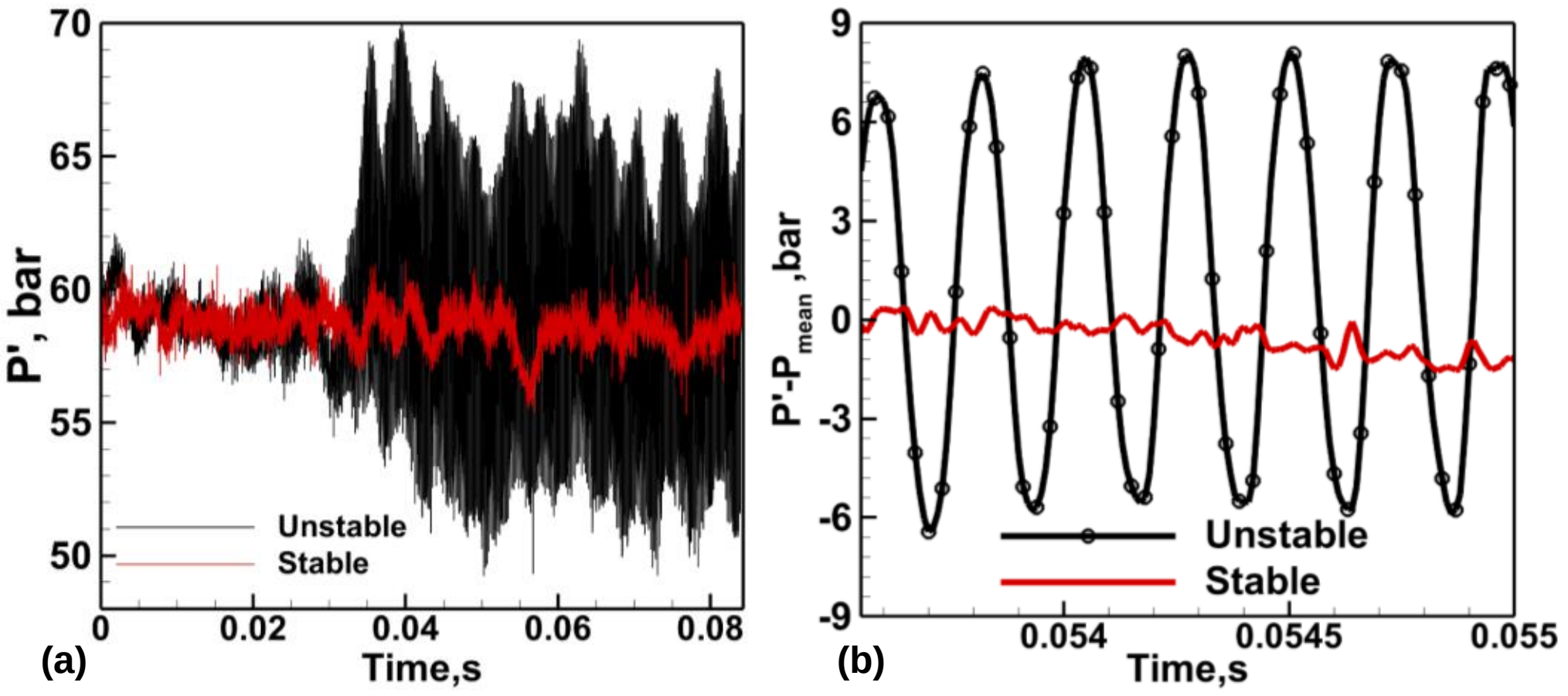

**FIG.7.** Comparison of pressure fluctuation: (a) stable and unstable cases, (b) zoomed view

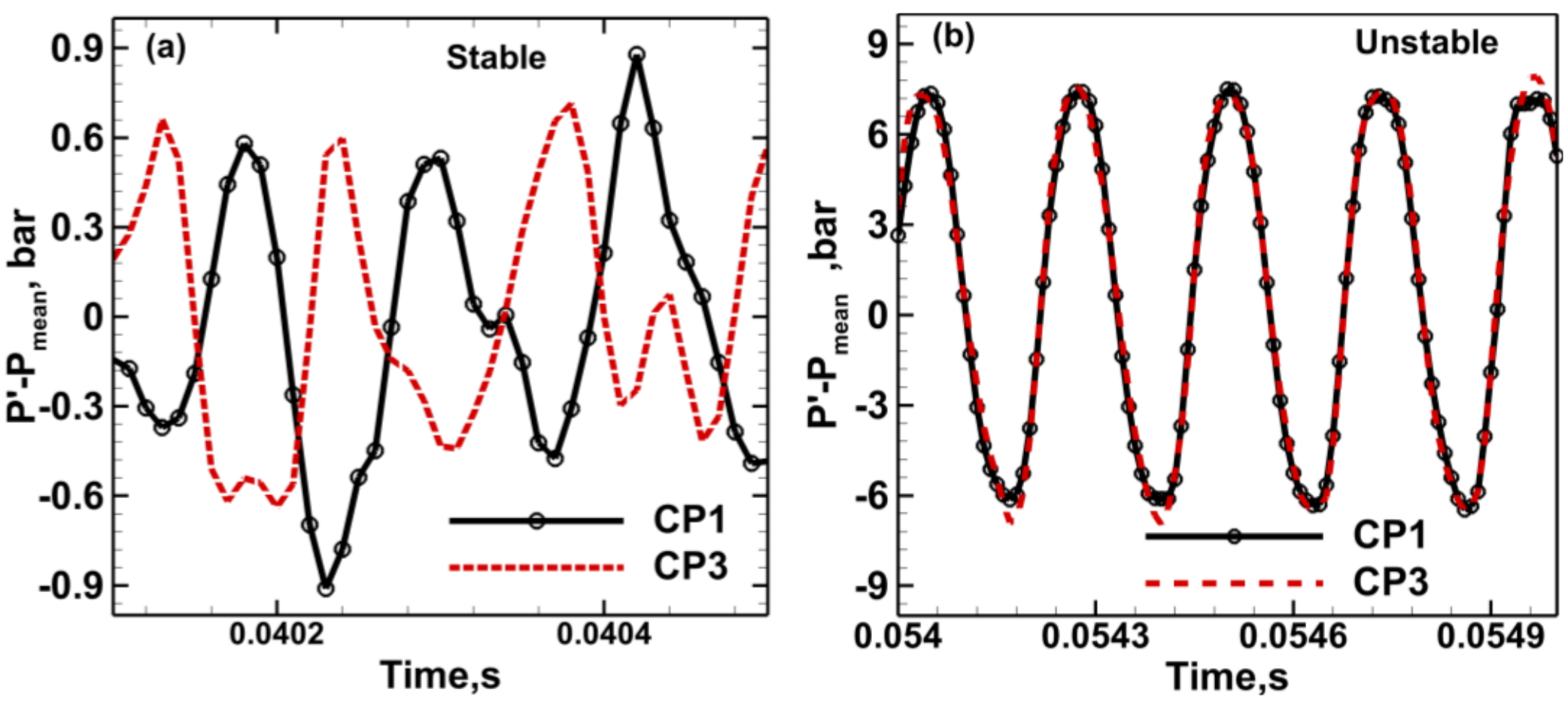

**FIG.8.** Top-bottom probe comparison for (a) stable, (b) unstable case

Figure 8(a) and (b) compare the CP1 (top) and CP3 (bottom) probes for stable and unstable cases within a smaller time window. Figure 8(a) shows an out-of-phase pressure variation at the near-wall probes, indicating the development of a tangential standing wave in the combustor. Although this out-of-phase pressure variation suggests a typical tangential standing wave mode, the fluctuation amplitude remains very low in the stable case, as shown in Figure 3(b). The time window displays 2.5 cycles of the pressure wave, corresponding to an acoustic frequency close to 8000 Hz. In contrast, the unstable case in Figure 8(b) shows an in-phase pressure wave, indicating the transverse mode's absence or very low amplitude in limit cycle oscillations. The in-phase variation at the top and bottom probes points to the onset of longitudinal wave movement inside the combustor. Figure 8 illustrates pressure fluctuation variations at the CP1 and CP7 probe locations to analyze the unstable case further. The out-of-phase pressure variations at the near-injector (CP1) and near-throat (CP7) positions indicate the onset of a longitudinal acoustic wave in the combustor, with a cycle frequency close to 4.3 kHz. The near-injector (CP1) and near-throat (CP3) positions represent two anti-nodal points in the combustor. The stable case shows no evidence of longitudinal wave pressure activity. In contrast, all unstable cases exhibit the same type of instability, characterized by longitudinal wave movement within the combustor. This raw pressure comparison reveals the nature of acoustic activity in stable and unstable cases, further explored through spectral and dynamic mode decomposition analysis in the next section. Before conducting spectral analysis, unsteady statistics are examined using cycle analysis for the unstable case to understand the instantaneous variation of parameters over one cycle.

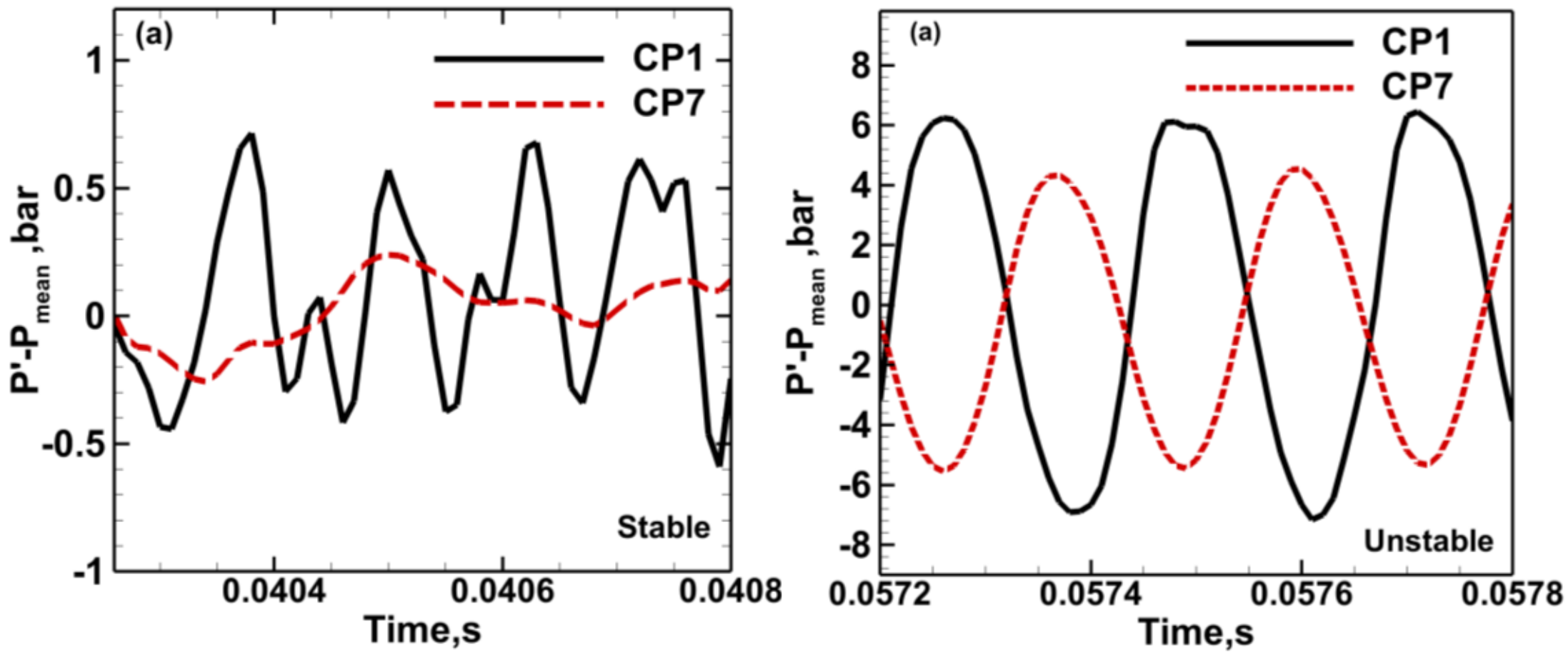

**FIG.9.** Start and end probe comparison: (a) stable, (b) unstable case

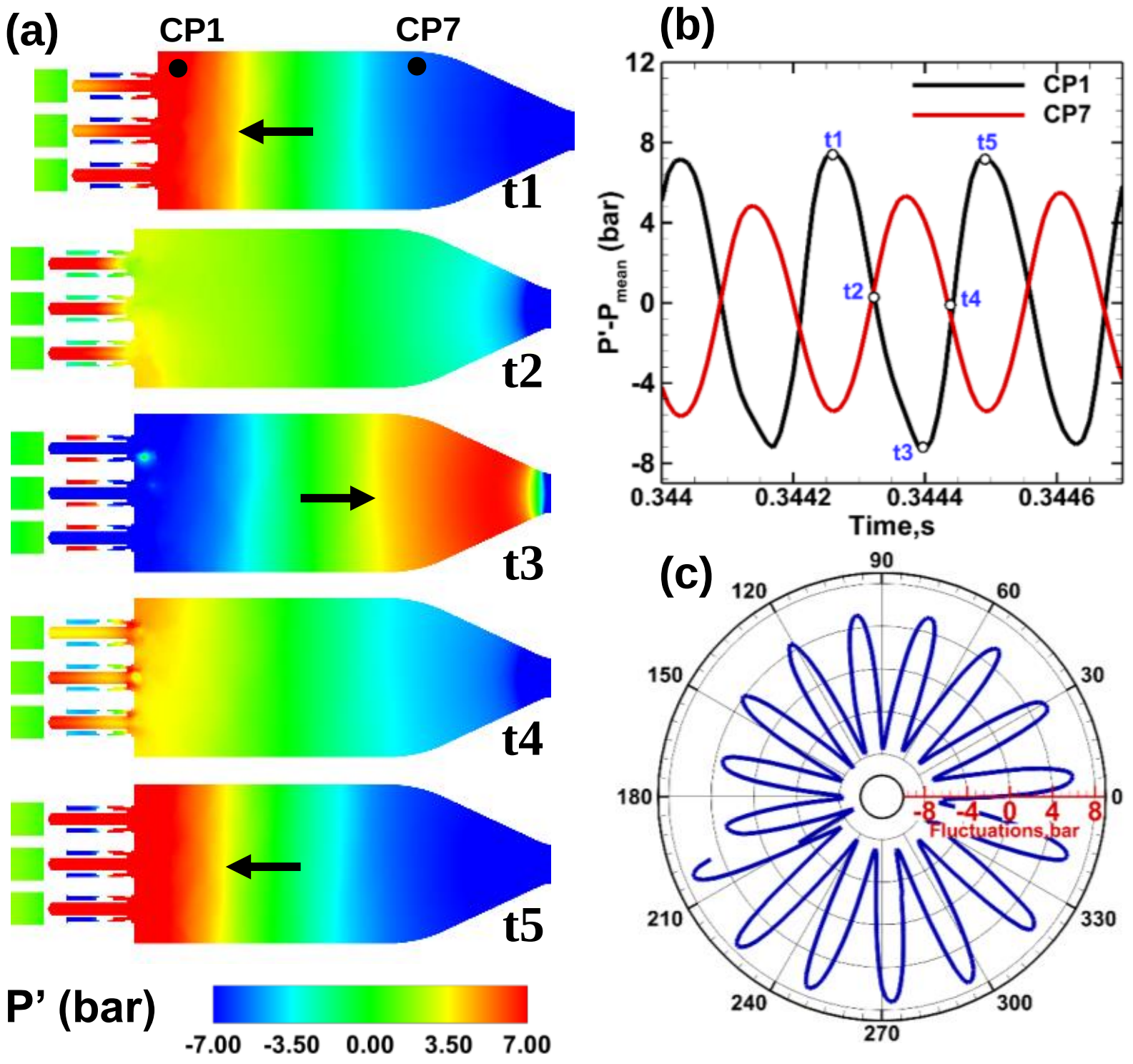

**FIG.10.** (a) Temporal variation of pressure, (b) probe variation, and (c) a polar sketch of pressure in an unstable case

Cycle analysis is conducted to understand the instantaneous variation of parameters at different time intervals over one acoustic cycle. Figure 10(a) presents a pictorial view of pressure fluctuation in the combustor at different time instants during an instability cycle. It illustrates wave movement in pressure contours at selected time instants t1 to t5, which represent the total time duration of one acoustic cycle. The arrows marked on the pressure contours highlight the movement of the high-pressure wave from the start to the end of the combustor and its return to the head end. This instantaneous pressure variation exhibits the characteristics of a typical longitudinal acoustic wave in the cavity, with alternating high- and low-pressure regions at fixed locations over time. Figure 10(b) shows the temporal variation of pressure as depicted in Figure 9(a), based on pressure variations at probe locations CP1 and CP7, as marked in Figure 10(a). It reveals a sinusoidal variation of pressure at both CP1 and CP7 over one acoustic cycle from time instant from time instant t1 to t5, but in opposite phases to each other. This variation confirms the formation of alternating high-pressure and low-pressure regions in the combustor under unstable conditions. Cycle analysis reveals the onset of longitudinal acoustic waves in the unstable case, with a high-pressure zone periodically forming near the injector faceplate, which can influence and sustain the instability. Figure 10(c) displays the acoustic wave movement on a polar chart, illustrating the pressure variation at a fixed point in the combustor during instability. The chart highlights alternating high- and low-pressure regions, represented as distinct peaks and troughs on the polar plot. These variations correspond to the longitudinal acoustic wave behavior in the combustor, showing periodic pressure fluctuations over time. The radial distribution reflects the magnitude of pressure at the given location, while the angular progression captures the temporal evolution within acoustic cycles. This visualization emphasizes the cyclic nature of pressure oscillations and the self-sustained nature of instability.

## B. Spectral Analysis

To further examine the frequency content across all simulated cases, a spectral analysis of the CP1 probe is conducted. Figure 11 presents an FFT comparison of each case's CP1 probe pressure spectrum, organized sequentially from the nominal 233 K down to the minimum injection temperature of 200 K. The top panel shows spectral plots for the stable cases, while the bottom panel compares low-temperature unstable simulations. In the bottom row, distinct high-amplitude peaks are observed at 4.3 kHz and harmonics at 8.6 kHz and 12.97 kHz for all unstable cases below 217 K. In contrast, the injection temperature cases from 233 K to 219 K do not exhibit these frequency peaks; instead, they show consistent peaks near 3.9 kHz and 7.9 kHz. These modes in the stable cases are characterized by very low-amplitude fluctuations, as noted in the raw pressure traces earlier. The dominant peaks in the FFT spectra below 219 K correspond to resonant frequency components associated with longitudinal wave movement in the combustor. These frequency tones are absent in stable cases.

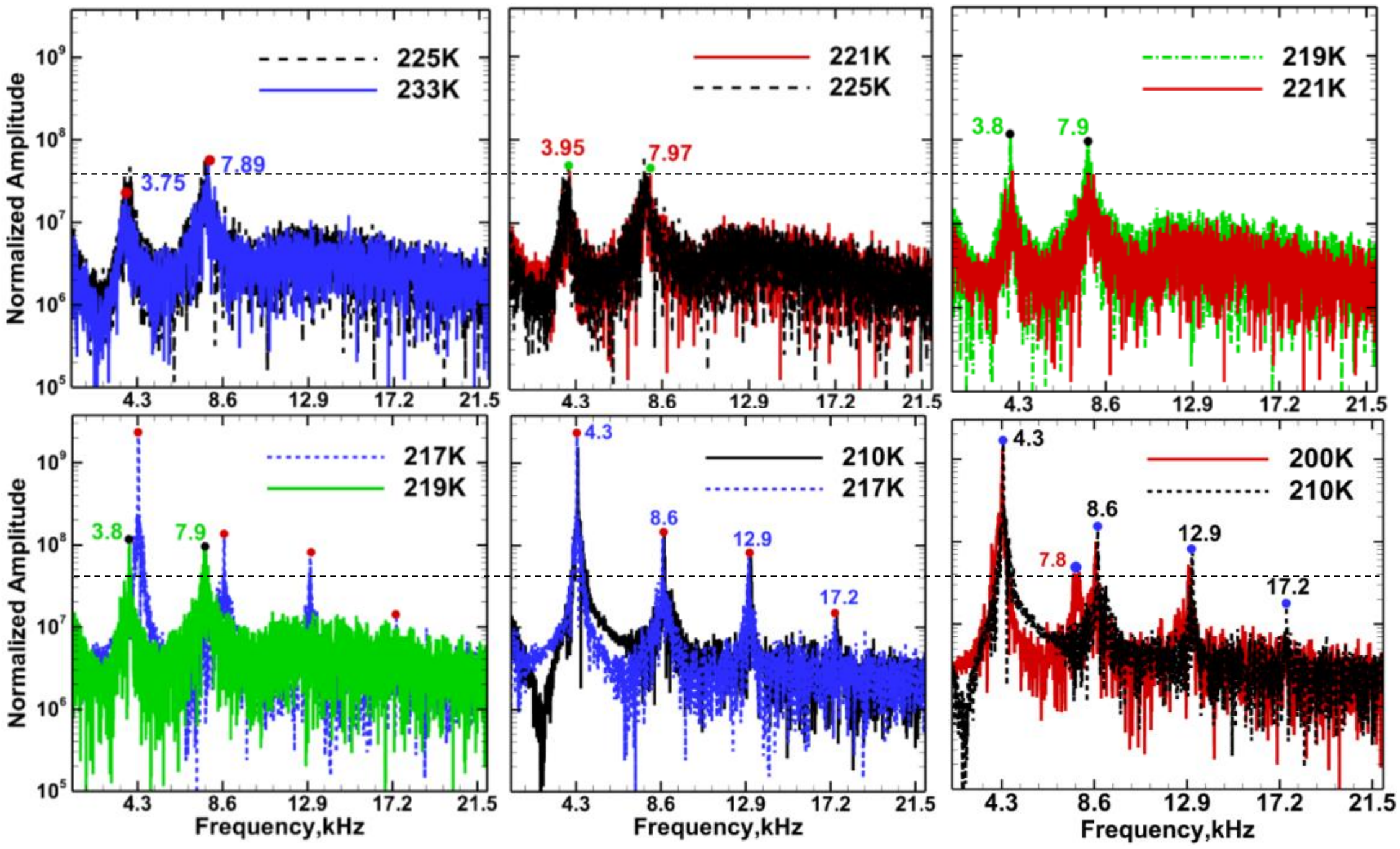

**FIG.11.** FFT comparison of all temperature cases in sequence

Figure 12 presents a three-dimensional view of spectral variation for all temperature cases. The 3.9 kHz frequency observed in the FFT spectra for all stable cases is identified as the first longitudinal (1L) resonant mode, closely matching the analytical value calculated at a constant equilibrium sound speed. An analytical estimate on acoustic mode frequencies (*f*) using a sound speed value of 1200m/s is calculated from NASA equilibrium solver at a chamber pressure of 62 bar and oxygen to fuel mass ratio (O/F) of 3.4. The resonant frequencies are calculated via:

$$f_{m,n.q} = \frac{c}{2\pi}\sqrt{\frac{\beta_{m,n}^2}{R_c^2} + \frac{q^2\pi^2}{L_c^2}} \tag{21}$$

where c denotes the speed of sound in the chamber, $R_c$ and $L_c$ represent the radius and length of the combustor, $\beta_{m,n}$ are roots of the Bessel function, and *m*, *n*, and *q* are the mode numbers. Further details on the analytical calculations for combustor acoustics can be referenced from our earlier work[20]. In contrast, in unstable cases, the 1L mode shifts to 4.3 kHz, which is attributed to the slightly higher sound speed in the combustor. Figure 13(a) provides a closer comparison of the spectral plots for both unstable and stable cases. It shows high-amplitude longitudinal wave modes 1L, 2L, and 3L, visible at 4.3 kHz, 8.6 kHz, and 12.97 kHz, respectively. In contrast, the stable case exhibits lower amplitude for the resonant modes at 3.9 kHz and 7.9 kHz. A comparison of the CP7 (combustor end) probe FFT, as shown in Figure 13(b), also reveals a similar trend, with dominant longitudinal modes exhibiting high amplitude in the unstable case, whereas the stable case displays low-amplitude peaks. It is observed that all unstable cases exhibit similar excitation of longitudinal acoustic modes in the combustor.

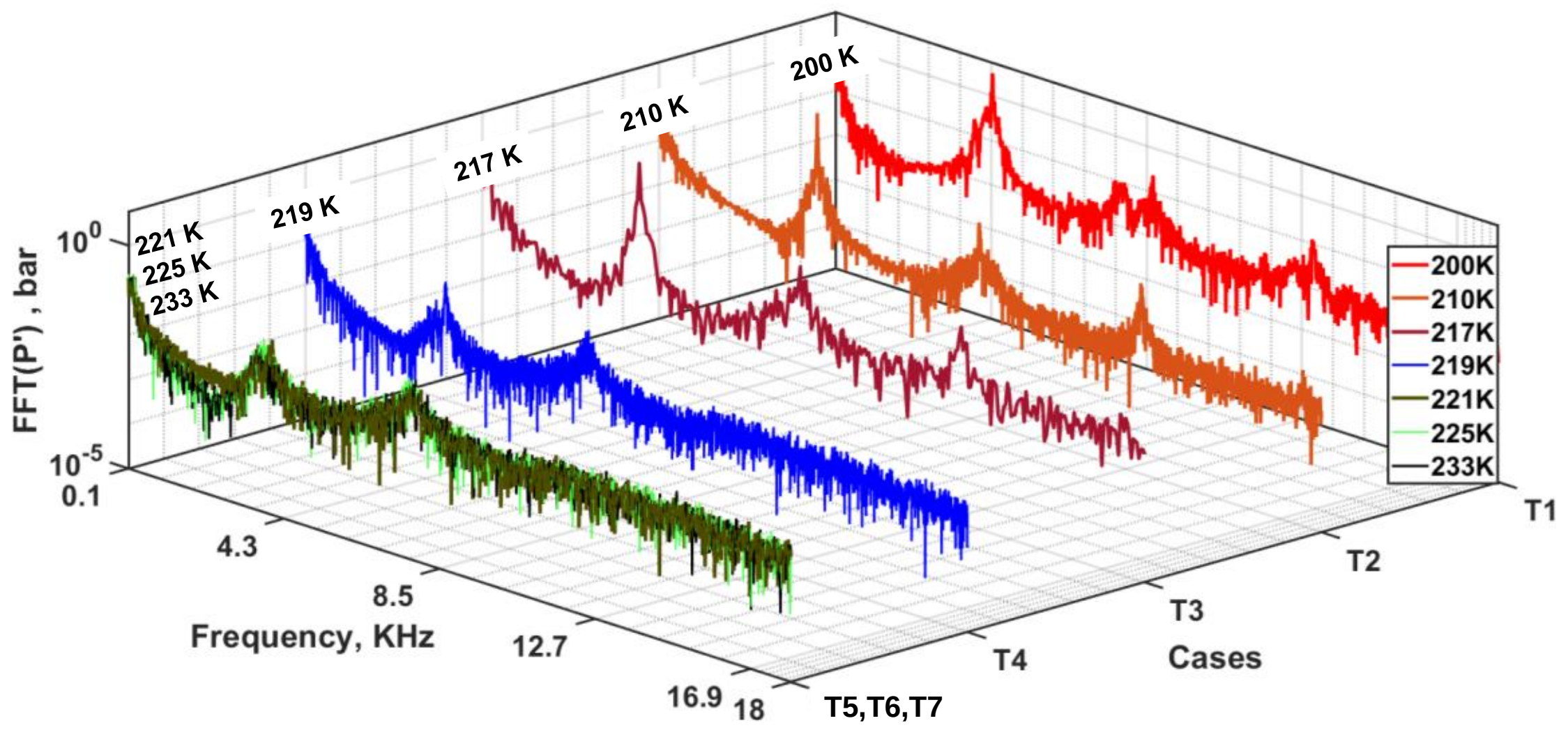

**FIG.12.** 3D spectral plot for all temperature cases

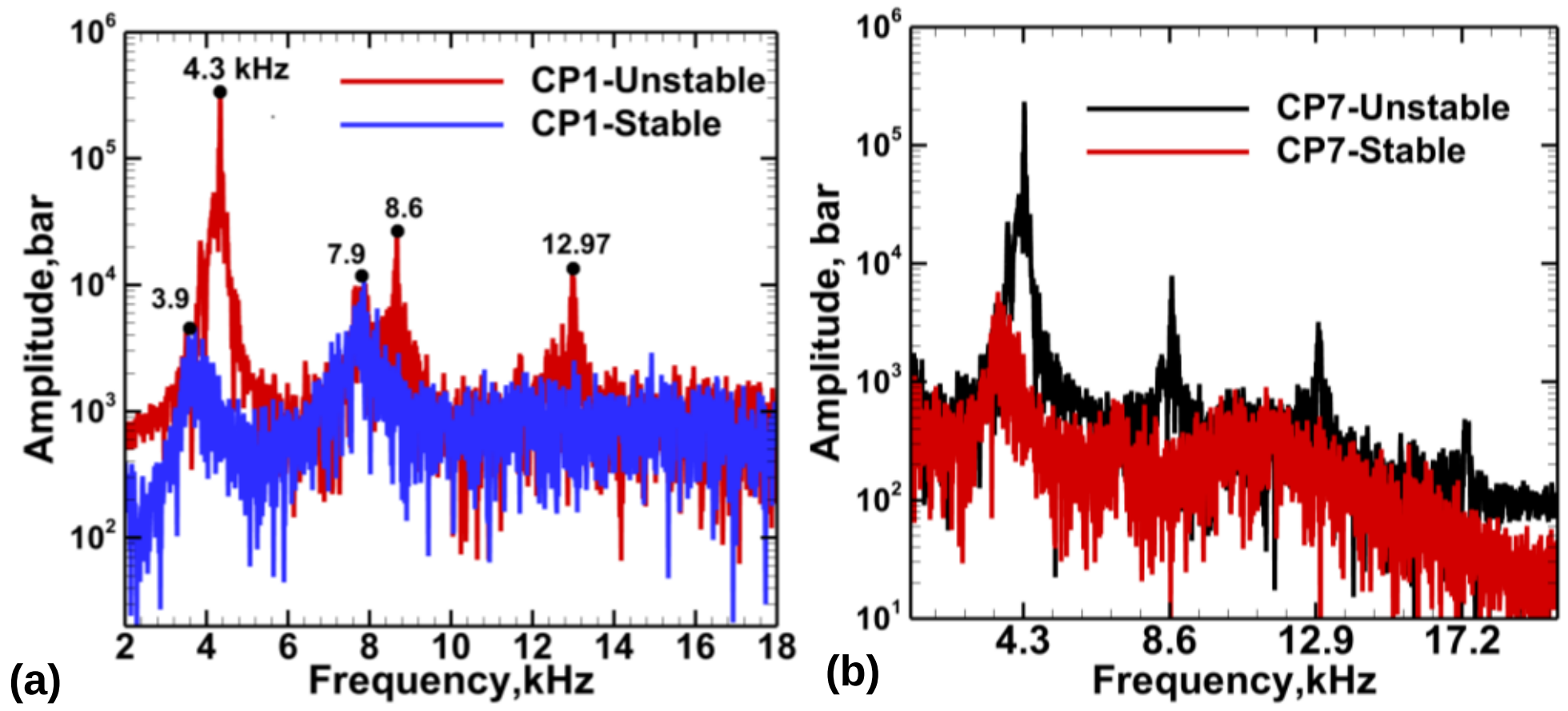

**FIG.13.** FFT comparison in unstable and stable cases: (a) CP1, (b) CP7 probe

Figure 14 compares various combustor probes in the unstable case, highlighting major resonant modes in the CP1, CP2, and CP3 spectra. While dominant longitudinal modes are evident, the analysis also reveals a 7.9 kHz frequency tone in the near-wall probes (CP1-top and CP3-bottom). Notably, the 1T mode frequency is absent in the center probe (CP2), aligning with the known behavior that transverse modes develop close to the injector faceplate and combustor walls. The absence of the 7.9 kHz peak at the center confirms the presence of the first tangential (1T) mode near the combustor walls. However, the amplitude of the first longitudinal (1L) mode and its harmonics (2L) is significantly higher than that of the tangential (1T) mode. It is observed that the 1T mode peak is visible only in the unstable case corresponding to a 200 K injection temperature and is not observed in other unstable cases at 210 K and 217 K, which predominantly show 1L mode excitation. The stable cases consistently show low amplitude peaks at 7.9 kHz. Figure 14(b) compares combustor end probes, showing high amplitude longitudinal mode peaks. It shows the apparent absence of 1T mode frequency at this location compared to the near injector spectrum shown in Figure 14(a).

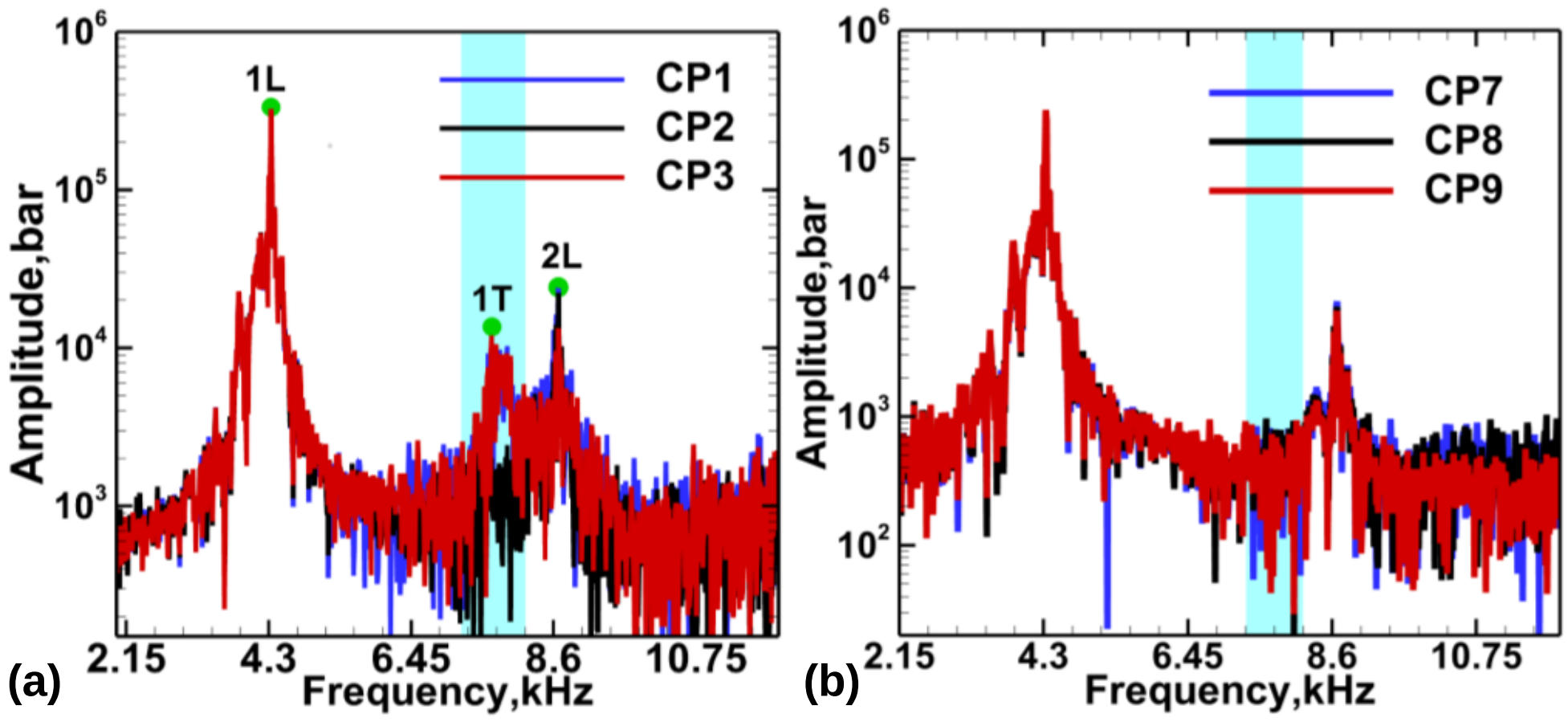

**FIG.14.** FFT comparison in unstable simulation: (a) start probes, (b) end probes

To further verify the evolution of the 1T mode in the combustor, a separate comparison of the center probes, CP2 and CP8, is conducted for both stable and unstable simulations. Figure 15 compares the frequency spectrum for center probes CP2 and CP8 in stable and unstable cases. It is observed that the CP2 spectrum for the unstable case shows dominant peaks for the 1L mode and its harmonics, with a noticeable absence of the 1T mode peak, as seen in the near-wall probes. In the unstable case, the CP8 probe displays four harmonics of the 1L mode, which are not present at the CP2 location in either the stable or unstable spectra. This difference is attributed to the location of the CP8 probe, which is unaffected by transverse wave propagation and thus clearly displays all triggered longitudinal modes. This comparison confirms that the center probes at both the combustor's start (CP2) and end (CP8) for both stable and unstable cases do not exhibit the 1T transverse mode at 7.9 kHz.

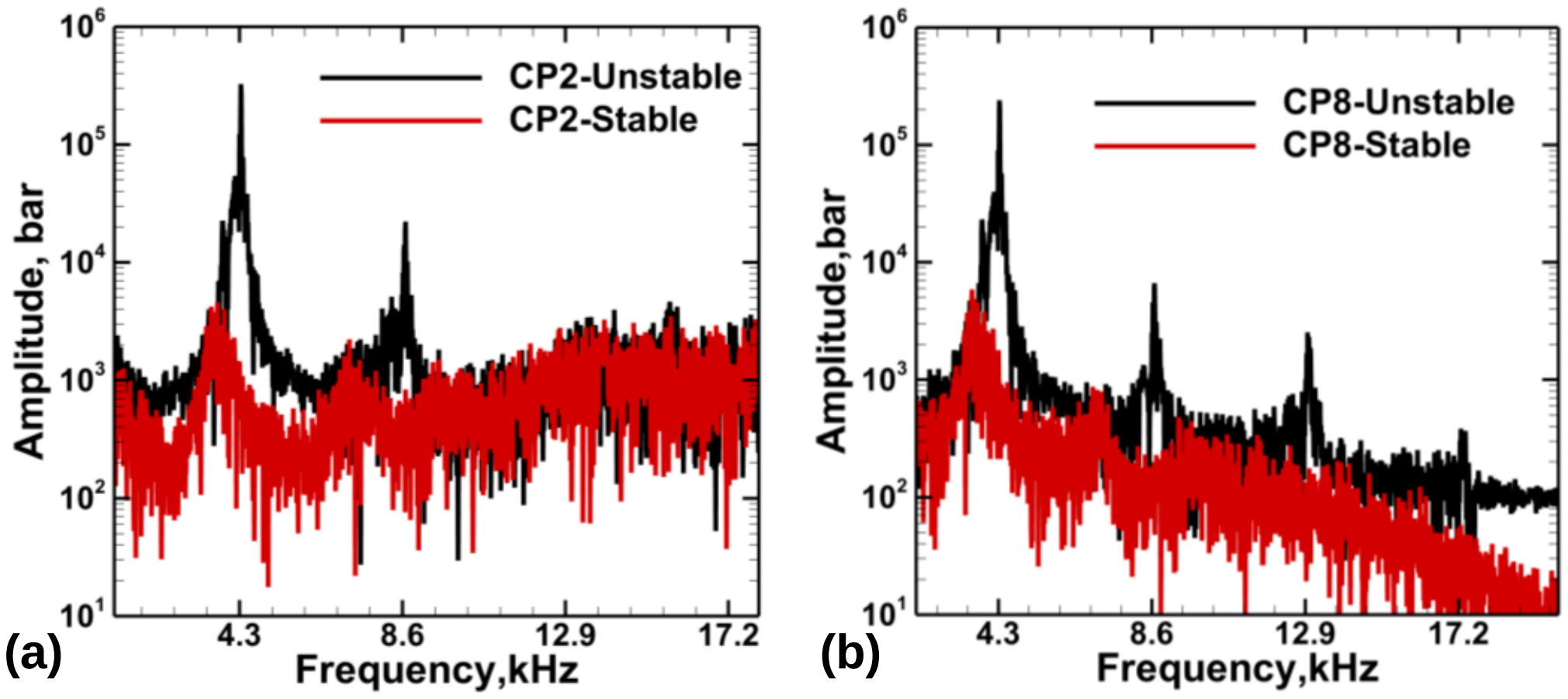

**FIG.15.** FFT spectrum of center probes for stable and unstable simulations: (a) CP2, (b) CP8

This spectral analysis reveals the excited modes in unstable cases, highlighting the onset of high-amplitude longitudinal acoustic waves when the fuel injection temperature drops below 219 K. Our study indicates that the 1L mode at 4.3 kHz and its harmonics possess the highest acoustic energy in unstable cases. The 1T mode at 7.9 kHz consistently displays a low-amplitude peak in stable cases. The lower-frequency longitudinal modes may require less energy to excite than higher-frequency transverse modes. In unstable cases, the available acoustic energy excites the combustor's 1L mode, whereas the 1T mode at a higher frequency may require a more extensive energy threshold to sustain resonance.

### C. Dynamic Mode Decomposition Analysis

Dynamic Mode Decomposition (DMD) is a data-driven technique to extract spatio-temporal coherent structures from time-resolved data to analyze nonlinear dynamical systems. The modern DMD algorithm, introduced by Schmid[45], applies to numerical and experimental data. It decomposes complex systems into dynamic modes with distinct growth rates and frequencies, represented by eigenvalues and eigenvectors, to identify coherent structures. Huang et al.[46] applied this method to analyze periodic combustion instabilities in gas turbines and model rocket combustors (CVRC). More recently, we[33] investigated injector-coupled combustion instabilities in a multi-element combustor using the same methodology. In this study, a parallel algorithm combining QR factorization and singular value decomposition (SVD), as proposed by Sayadi et al.[47], is employed to perform the DMD analysis. The details of the DMD methodology can be found in our previous work[33]. The DMD methodology used in this study has been effectively applied to different test cases, as reported in earlier works[42,48–50]. This study uses a DMD analysis of the combustor's stable and unstable temperature simulations to compare the spatial modes. The analysis aims to identify characteristic differences in the combustor mode shapes for stable and unstable operations. As noted earlier, all unstable cases exhibit similar spectral features; therefore, DMD analysis is presented here for a typical unstable case. We used 400 pressure snapshots from the axial plane, gathered at a sampling rate of 100 kHz, corresponding to over 150 acoustic cycles for the 1L mode. Figure 15 displays the spatial mode shapes for both stable and unstable cases. The first five modes capture the most energy and are examined to gain insights into the coherent structures.

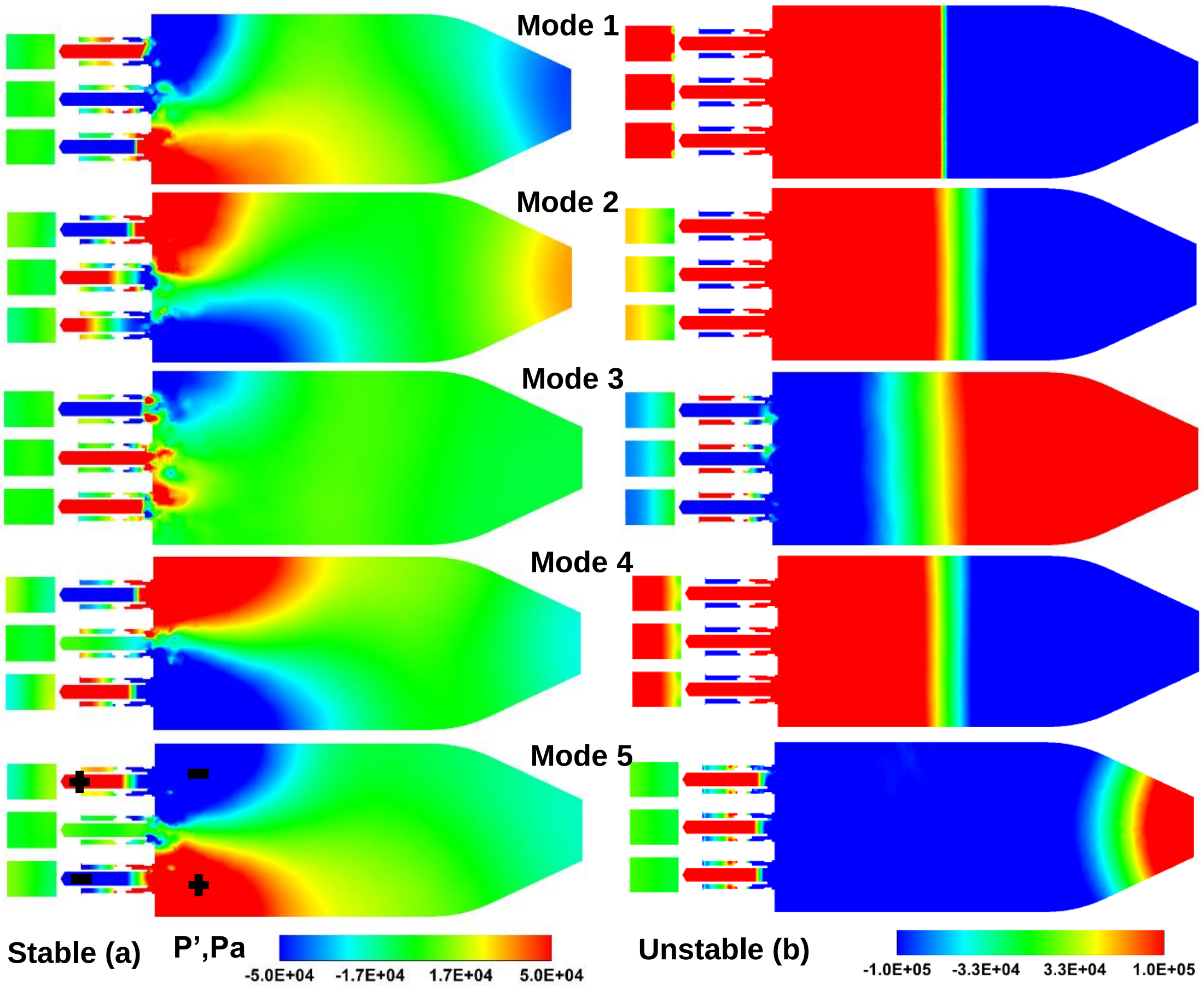

**FIG.16.** DMD modes for (a) stable and (b) unstable cases

Figure 16(a) presents the DMD modes for the stable case in the first column. The stable case shows typical tangential wave modes in the combustor, with transverse spatial modes in the chamber and a longitudinal standing wave visible in the injector section. This spatial representation also highlights the influence of combustor modes on the evolution of acoustic waves within the injector region. The DMD modes also correspond to a frequency of 7.9 kHz, which matches well with the 1T mode frequency identified in the spectral analysis. Figure 16(b) presents the spatial modes for the unstable case. The mode shape clearly illustrates the longitudinal wave movement within the combustor and injector sections. Mode 1 represents the fundamental mode in the chamber, characterized by the 1L wave distribution. It captures the dominant 1L mode at 4.3 kHz, followed by the 2L mode in mode 5. The DMD analysis also confirms the excitation of the 1L mode, as observed in the spectral analysis of the unstable cases. The mode shapes for this unstable case indicate that the acoustic waves in the LOx paths of the injectors vary in sync with the combustor, while the methane entry into the injectors is not in phase with the combustor modes. This out-of-phase variation can modulate methane flow, potentially contributing to the self-sustaining nature of instability. The injector dynamics driving instability in low-temperature cases are illustrated in the next section.

## D. Injector Dynamics

Injector dynamics are examined through pressure time histories and spectral analysis of data from injector probes. Pressure and other parameters are collected along the injector's oxygen and methane paths at various probe locations. Figure 4 shows IP (oxygen path) and IM (methane path) denote injector probes. Multiple probes gather unsteady statistics and reveal how injector dynamics influence combustor stability. This section provides a comparative analysis of injector dynamics for stable and unstable cases. Figure 17 presents the pressure patterns of injector probes in the oxygen path for unstable (a) and stable (b) cases. In the unstable case, all probes show in-phase variation, whereas the stable case exhibits out-of-phase variation. The DMD modes pictorially illustrate this observation, showing intermittently varying higher and lower pressure zones in both the combustor and injector regions for stable cases. A significant difference in amplitude is observed between the two cases.

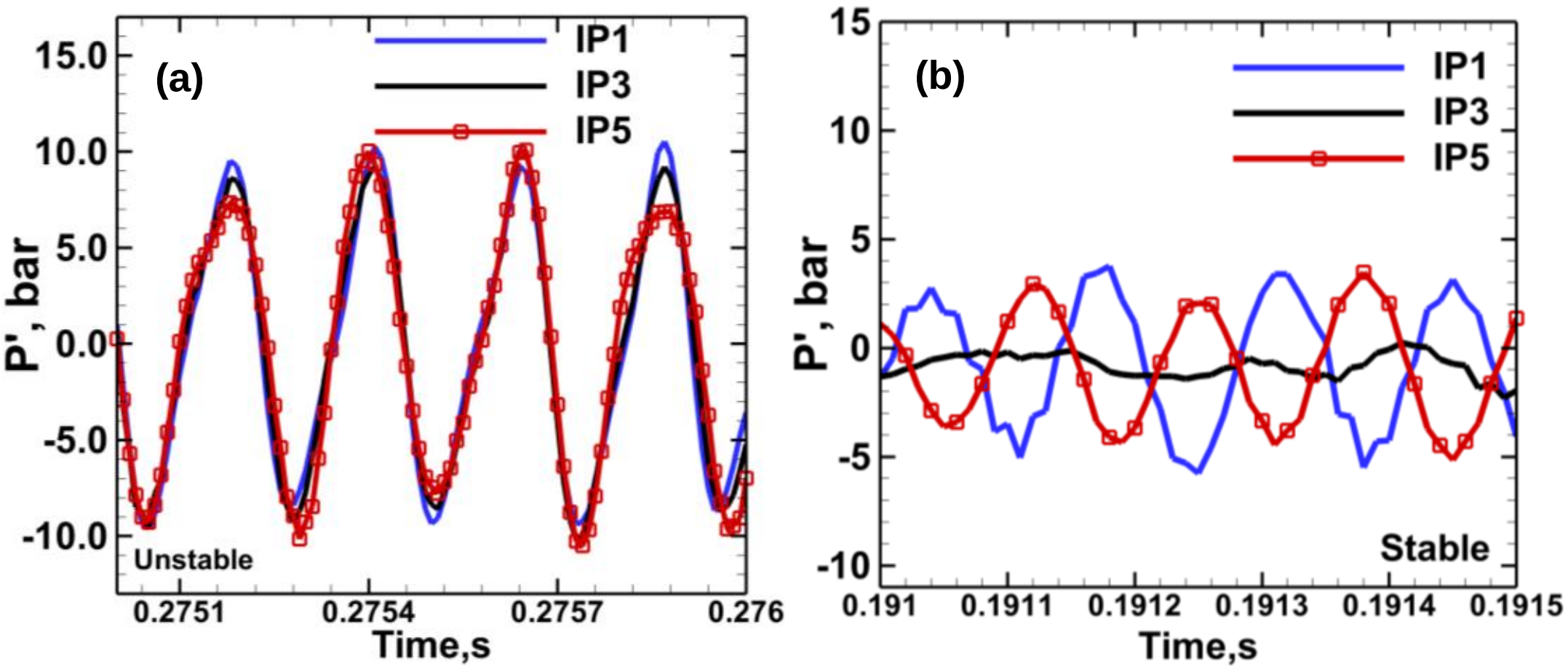

**FIG.17.** Pressure pattern in injector probes for (a) unstable and (b) stable cases

Figure 18 compares the pressure evolution in the injectors alongside the combustor. It also shows in-phase variation for all combustor and injector probes in unstable cases. In the stable case, lower injector pressure at IP1 corresponds to higher combustor pressure at the CP1 location. As mentioned earlier, this behavior is attributed to the tangential wave movement in the stable case. However, the amplitude of the tangential wave is lower than that of the longitudinal wave in unstable cases.

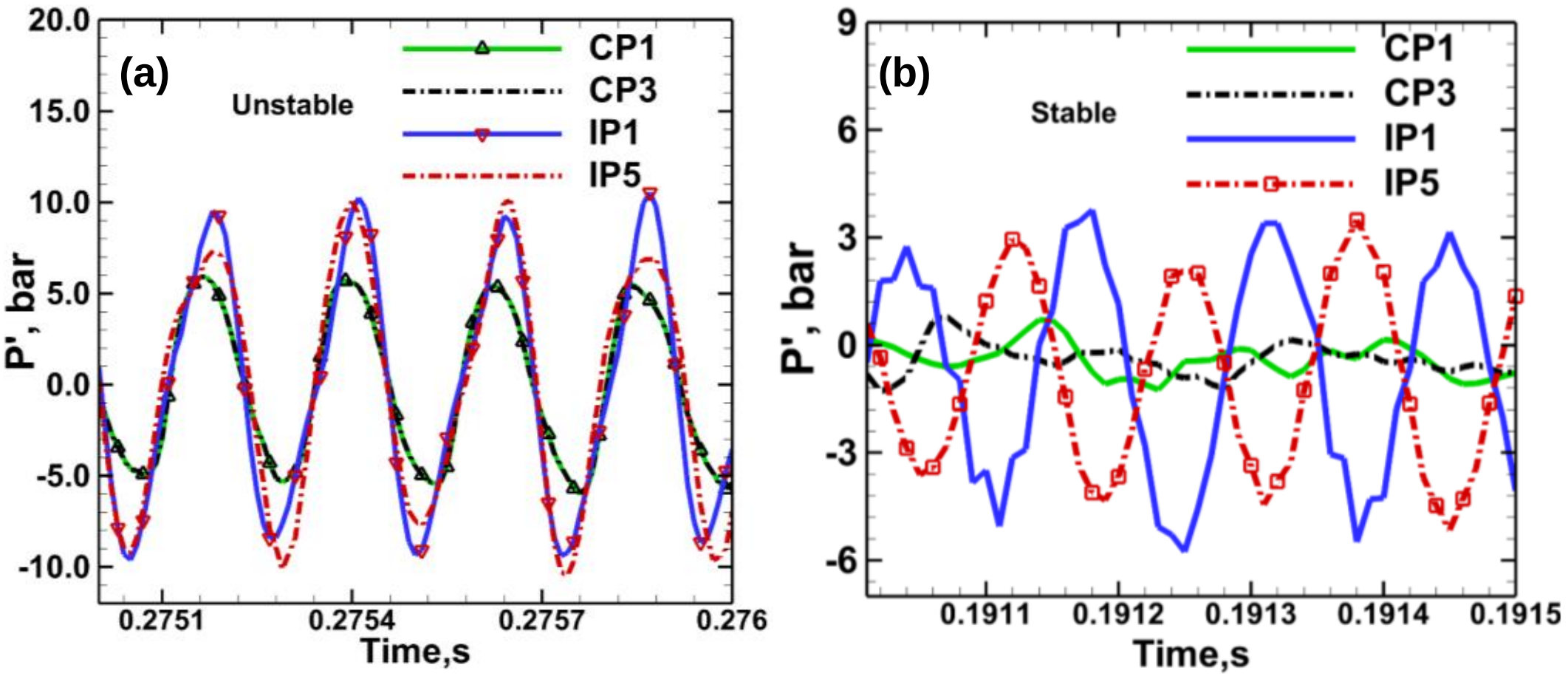

**FIG.18.** Pressure comparison of injector and combustor probes: (a) unstable, (b) stable cases

We now discuss the characteristic difference in oxygen path acoustics influenced by longitudinal and tangential wave movement. The in-phase variation observed in unstable cases and the out-of-phase variation in stable cases may be vital to understand the self-sustained nature of instability observed in low-temperature cases. We present a comparative assessment of injector modes in stable and unstable cases to examine the injector-associated combustion dynamics. Fast Fourier Transform (FFT) analysis is conducted on probes placed within the injector section to capture acoustic wave movement in the oxygen path of the injectors. Acoustically, the oxygen post can be modeled as a quarter-wave resonator, with the injector entry treated as a closed end and the exit as an open boundary. This configuration exhibits longitudinal modes with a natural frequency calculated as $f = c/4(L + \Delta L)$, where L is the injector length, c is the speed of sound, and ΔL is the correction factor, 0.6 times the oxygen post radius ($R_n$). The theoretical frequency of the first longitudinal mode (1L) at an oxygen sound speed of 850 m/s is 7450 Hz. Figure 19 shows spectral graphs for each injector point, indicating a peak frequency of approximately 7.5 kHz, closely aligning with the theoretically predicted longitudinal mode frequency. In the spectral plot for the unstable case, combustor resonant modes are visible, while the injector mode at 7.5 kHz appears at all IP1, IP3, and IP5 injector start locations. This injector mode peak is present in both stable and unstable cases. The spectral plots demonstrate the quarter-wave characteristics of the injector, with the amplitude of the 7.5 kHz peak frequency decreasing from the injector's entry to its exit, as indicated by the dotted line across all three injector locations. The amplitude falls from a maximum at the inlet probe (IP1) to a minimum at the near-exit probe (IP2), a pattern also observed in the other two injectors. Longitudinal acoustic waves dominate pressure oscillations within the injectors for stable and unstable simulations. This comparison shows that, in the unstable case, injector dynamics are influenced by the combustor's excited resonant modes, with the injector's oxygen path modulating in sync with the combustor and potentially leading to mass flow rate fluctuations. The following section presents the instability mechanism, illustrating the self-sustained nature of instability in low-temperature, unstable cases.

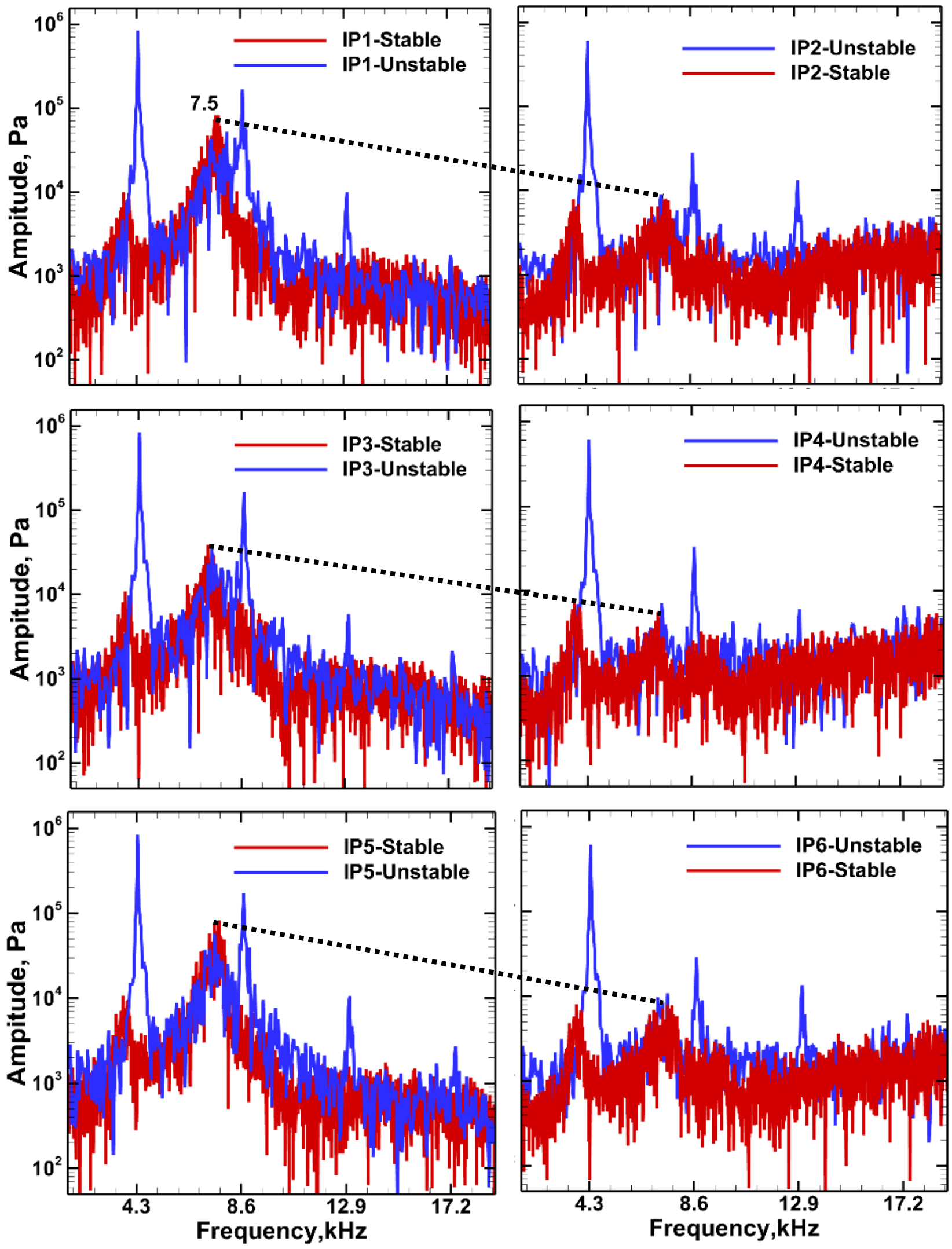

**FIG.19.** FFT spectrum of injector probes for stable and unstable simulations

## E. Comparison of Flow Features

Raw pressure oscillations reveal the onset of instability in low-temperature cases, which is confirmed by detailed spectral and dynamic mode decomposition analysis. Frequency analysis reveals that high-amplitude pressure oscillations occur in the first longitudinal mode of the combustor. The instability mechanism is further explored in the next section, which examines the potential delay in combustion caused by lower-temperature fuel injection. The higher potential for a feedback loop between unsteady pressure and heat release is discussed through the spatial variation of essential flow and flame parameters. Figure 20(a) displays the instantaneous pressure contour at a simulation time of 80 ms, showing the spatial pressure variation in both stable (top) and unstable (bottom) simulations. As previously discussed, the unstable case exhibits the formation of a high-pressure zone near the injector faceplate and a lower-pressure region at the opposite end, characteristic of longitudinal wave movement. Figure 20(b) compares the temperature variation for both cases, highlighting a compressed flame zone near the injector outlet in the unstable case compared to the stable case. This illustrates the impact of longitudinal wave movement on the flame front, which the incoming pressure wave can distort.

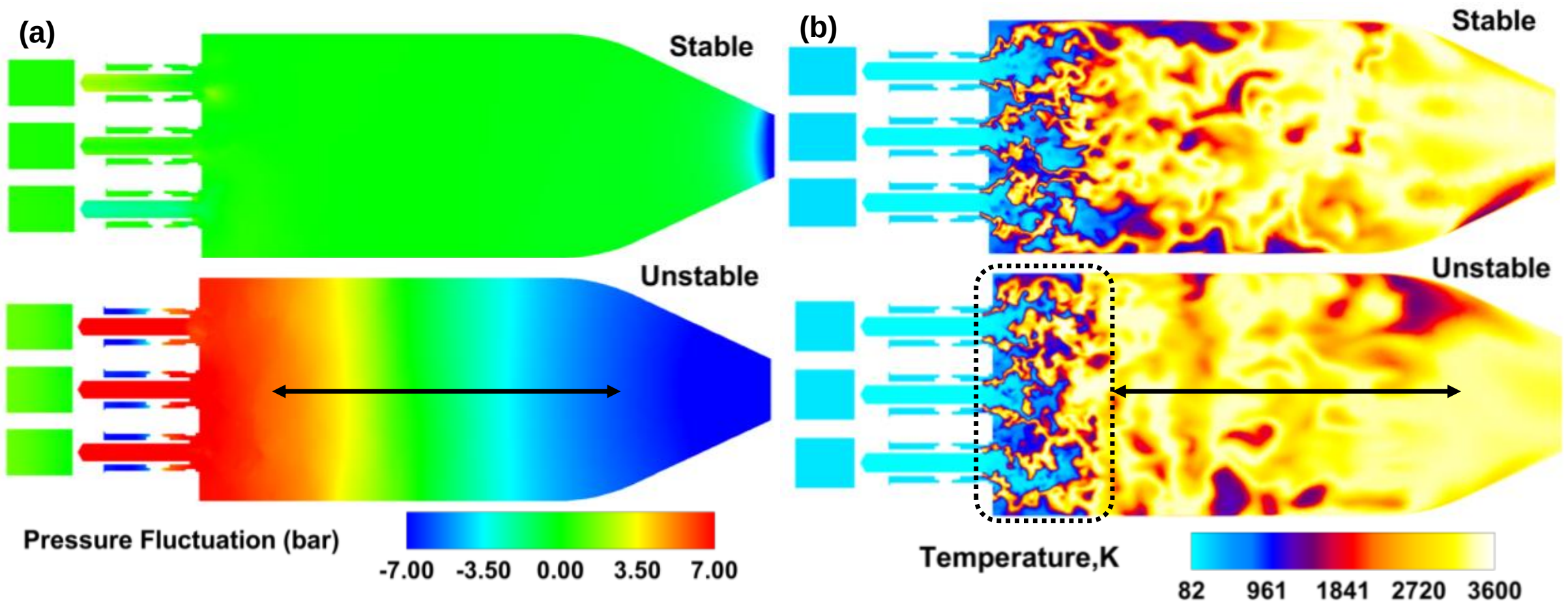

**FIG.20.** Comparison at cut plane: (a) instantaneous pressure and (b) temperature variation

We discuss the flame anchoring in stable and unstable simulations to determine the characteristic differences in both cases. Figure 20 displays streamline contours for both cases.  As described earlier, the swirl injectors used in this study promote rapid mixing, with the flame anchoring in the recess region of the oxygen post. The bi-directional swirl motion of methane and oxygen creates toroidal recirculation zones that aid in flame stabilization. Figure 21 shows streamlines emanating from the injector near the wall. In the unstable case (Figure 21(b)), a large recirculation zone, labeled as 1, appears near the upper wall, with two smaller counter-rotating bubbles (2 and 3) below it. In contrast, the stable case (Figure 21(a)) displays smaller recirculation zones. These corner recirculation zones continuously supply hot combustion products, aiding flame anchoring within the recess region. Under low-temperature conditions, larger recirculation zones can retain more fuel-rich mixtures, increasing combustion intensity in the shear layer and contributing to sustained combustion instability.

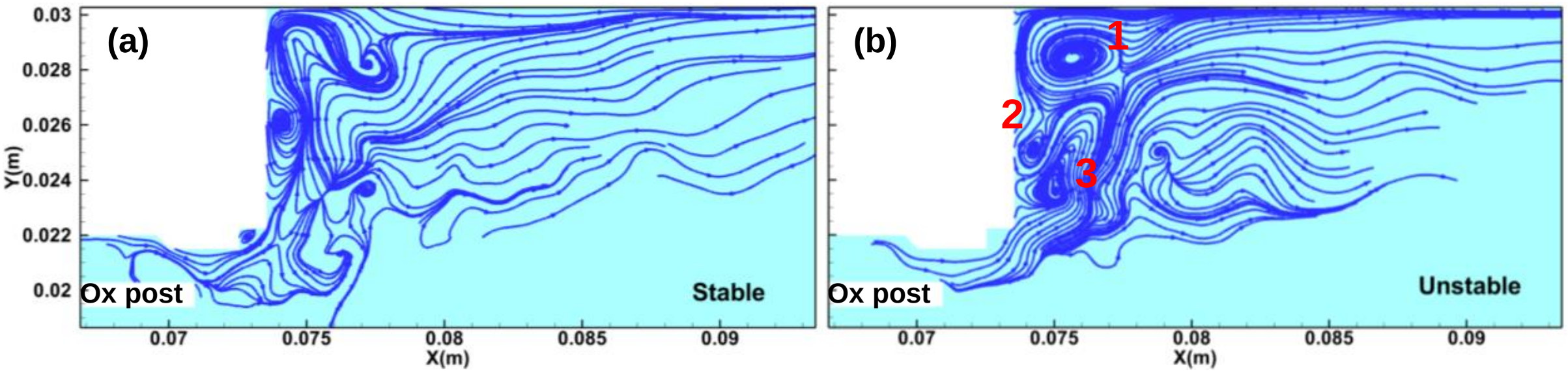

**FIG.21.** Streamline contours:  (a) stable and (b) unstable simulations

The self-sustained nature of the acoustic wave in unstable cases is further investigated by comparing flame parameters. The acoustic wave can compress the ejected methane and oxygen near the injector faceplate, potentially altering the chamber's local mixture ratio (O/F). Figure 22(a) compares the time-averaged mixture fraction at an axial plane, showing a higher methane concentration in the low-temperature unstable case, as indicated by the higher mixture fraction in the chamber region. The effect of the acoustic wave on the sound speed contour is also shown in Figure 22(b), which reveals a higher overall sound speed at the injector exit in the unstable case compared to the stable case. As discussed in the spectral analysis, the increased sound speed at this anti-node location results in a slightly higher resonant frequency.

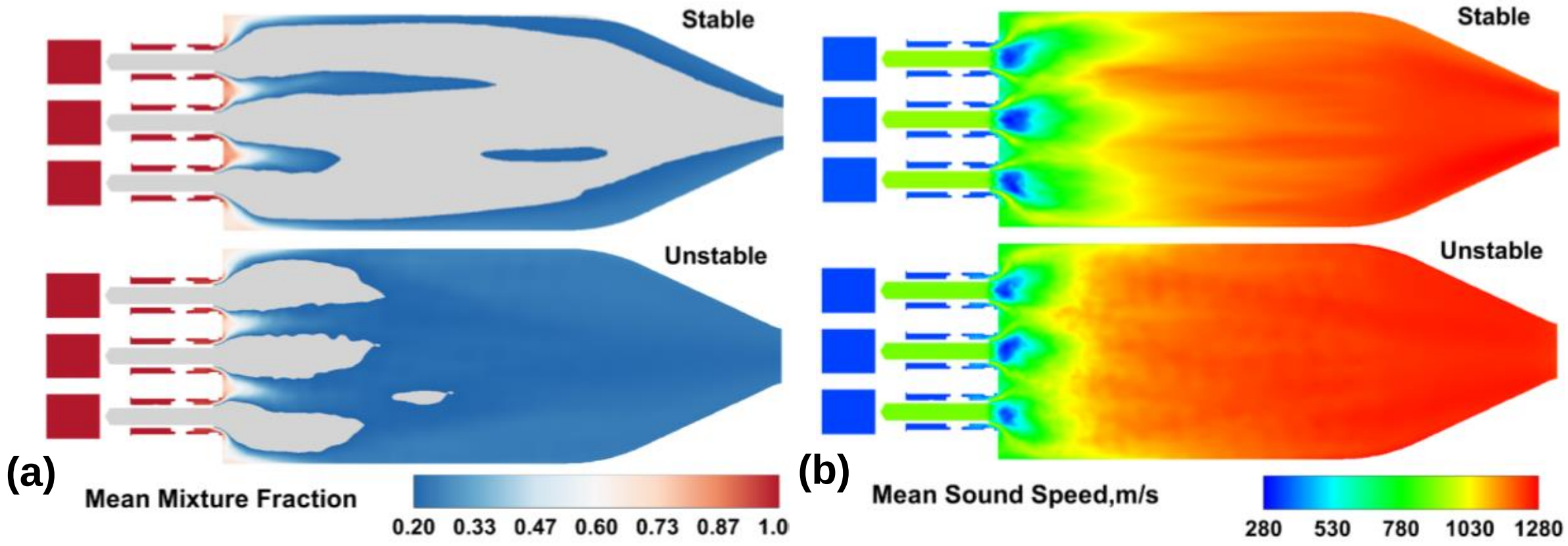

**FIG.22.** Mean Mixture fraction and mean sound speed: (a) stable and (b) unstable cases

We compared the mean vorticity and heat release rate in stable and unstable cases. Figure 23(a) compares mean axial vorticity production, revealing the instability mechanism through vorticity production in both stable and unstable cases. A higher magnitude of vorticity at the injector exit location is observed in the unstable case, which coincides with an increase in pressure at the head end location. Figure 23(b) compares mean heat release, showing a higher heat release near the injector faceplate in the unstable case. The higher heat release at the anti-node location triggers the onset of longitudinal wave movement and establishes a feedback loop. This comparison highlights the combined presence of higher vorticity and heat release near the combustor start location, which is crucial in sustaining this instability. Finite rate reactions play a crucial role in instability. The mean finite reaction rate for stable and unstable cases is compared to further elaborate on the flame features associated with unstable cases. Figure 24(a) compares the mean finite-rate source at the axial plane, showing a more considerable contribution of the finite-rate source term in the transport of the progress variable. It can be observed that the higher finite-rate source at the antinode location for longitudinal waves may trigger this instability mode in the combustor. This also highlights the appropriateness of FGM formulation, which only considers the effect of finite rate reactions in diffusion flames. Figure 24(b) displays the mean $H_2$ mass fraction. It reveals a higher concentration of $H_2$ in the combustor start region and globally for the unstable case compared to the stable case. The reaction zone cannot overcome heat loss at lower methane injection temperatures, allowing $H_2$ to persist without consumption. A higher concentration of $H_2$ indicates incomplete and erratic combustion, which can lead to temperature and pressure fluctuations in the combustor, potentially sustaining instability.

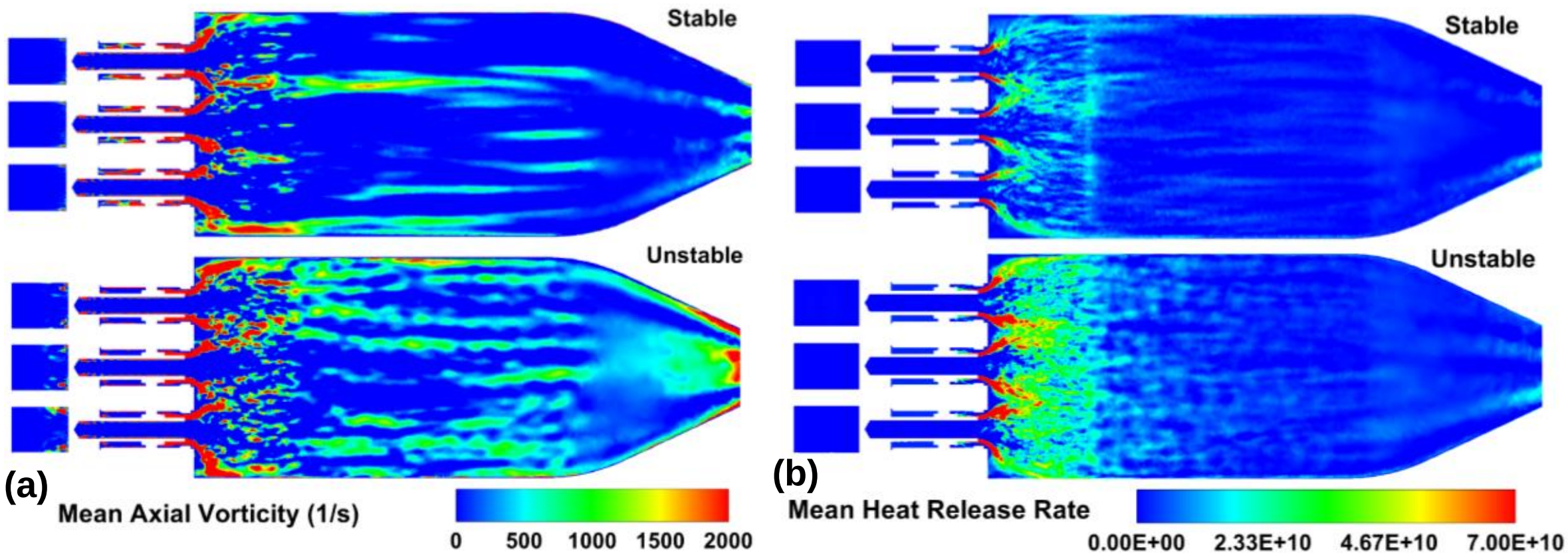

**FIG.23.** (a) Mean axial vorticity, (b) mean heat release contours for stable and unstable case

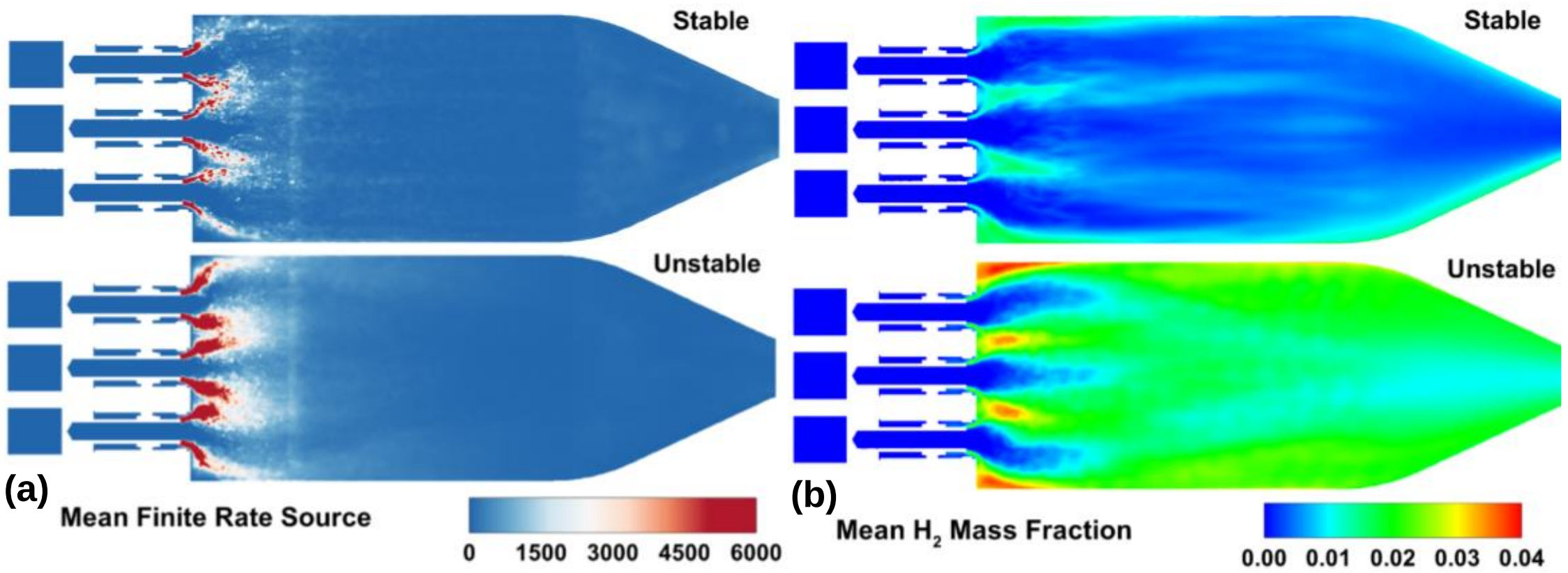

**FIG.24.** : (a) Mean finite rate source and (b) $H_2$ mass fraction contours for stable and unstable case

Our detailed comparison of flow and flame characteristics in stable and unstable cases illustrates the conditions that can lead to self-sustained instability at low temperatures. A feedback loop is established in low-temperature cases, driven by finite-rate effects, higher vorticity at the combustor's start location, and incomplete combustion, as indicated by higher $H_2$ concentrations.

## F. Insights into the Mechanisms of Instability

A comprehensive effort has been made to identify the mechanism responsible for triggering instability in low-temperature cases. This section explores the instability mechanism, emphasizing the possibility of a delay in combustion due to lower fuel injection temperature leading to a potential feedback loop between unsteady pressure and heat release. Probable instability mechanisms are discussed in detail through major spatial and spectral variation of parameters. Pressure fluctuations are widely recognized as a result of fluctuating heat release. The coupling between pressure and heat release is a core feature of combustion dynamics. Figure 25 illustrates the transition to instability in a low-temperature case. Here, S denotes stable, T represents transition, and US marks unstable. It displays a parallel variation of instantaneous pressure and heat release at the combustor axial plane. The stable state exhibits less intense heat release profiles, while the unstable state, post-transition, shows intense heat release near the injector exit. A clear transition from a stable state to an unstable condition for low-temperature LES is captured.

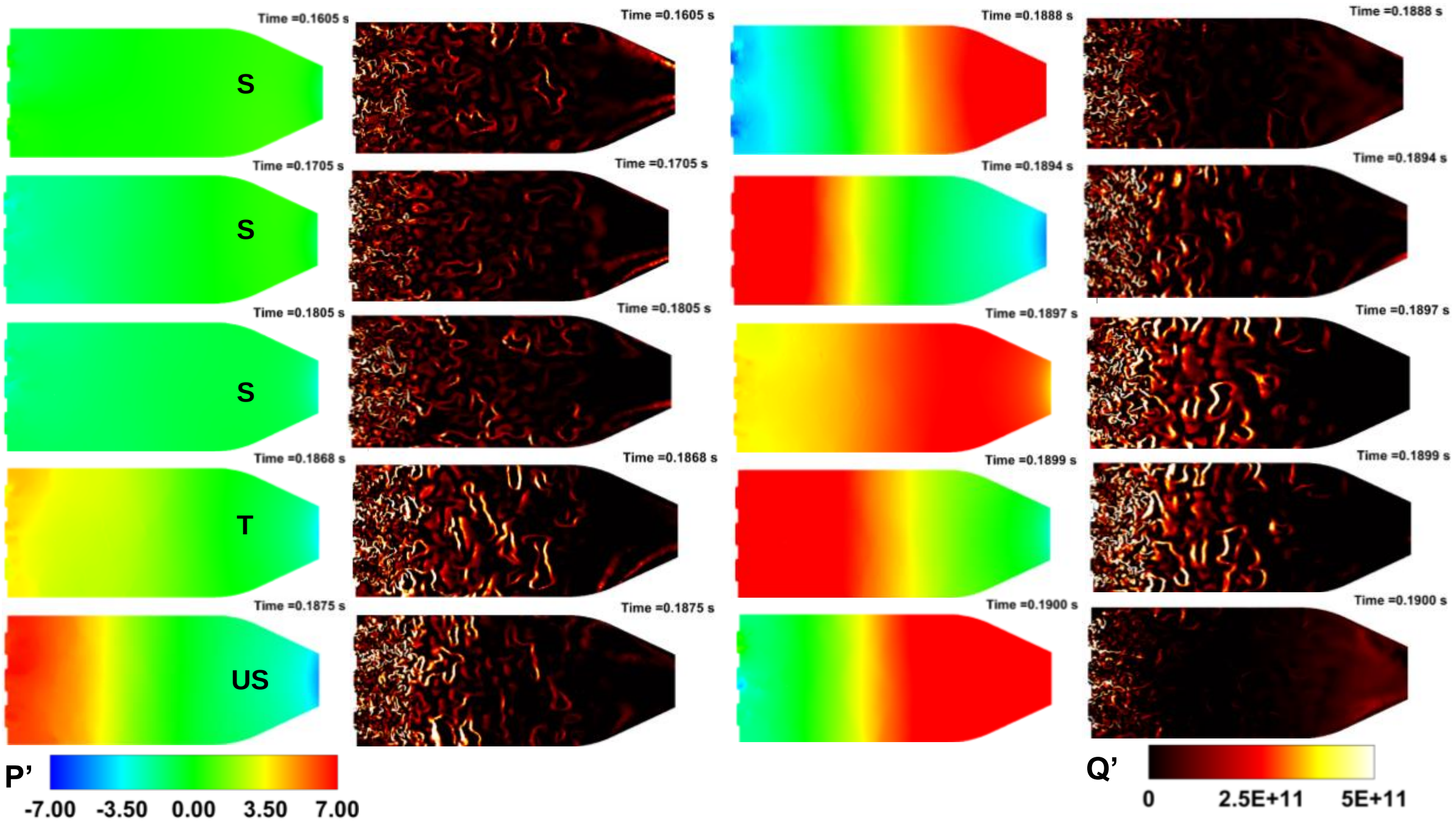

**FIG.25.** Instantaneous pressure (P') and heat release (Q') contours for transition to instability

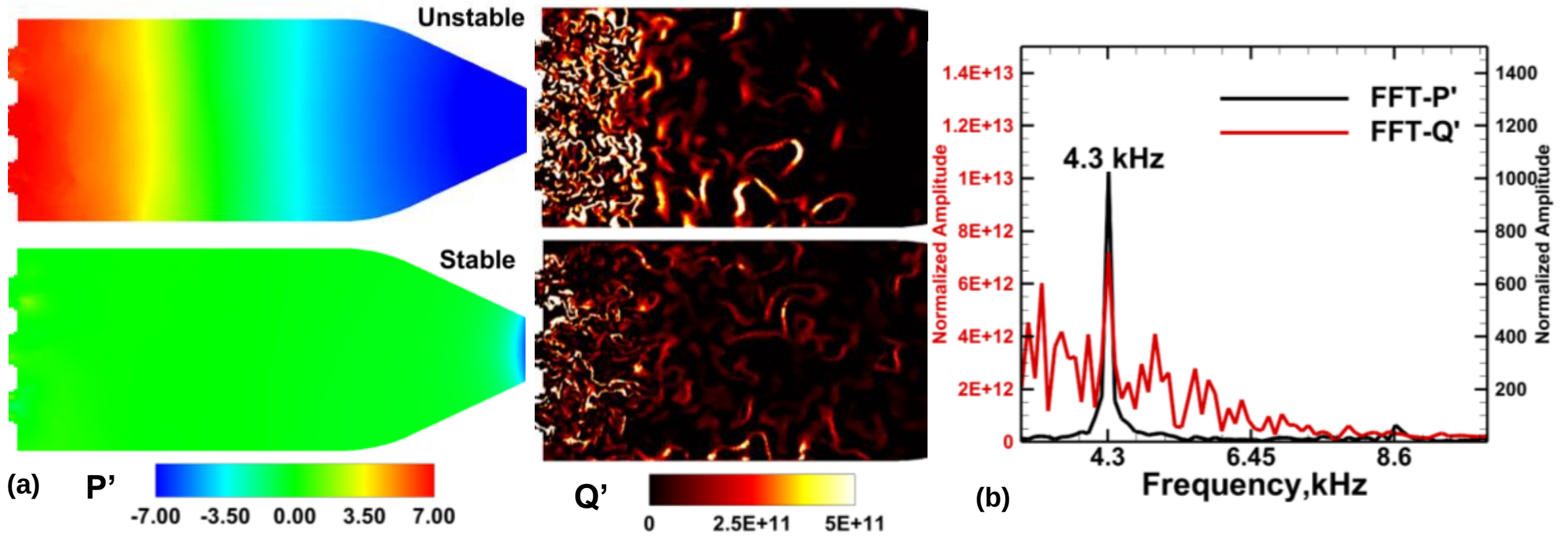

**FIG.26.** (a) Comparison of instantaneous pressure (P') and heat release (Q') for unstable and stable simulations at a particular time instant (b) FFT comparison for unstable case

Figure 26(a) provides a detailed view of pressure fluctuations and heat release rates for stable and unstable conditions at a specific time instant. It highlights a high heat release zone corresponding to high pressure near the injector faceplate at the same time instant. In the stable case, pressure variation is negligible, and no abrupt heat release is observed at the axial cut plane. In contrast, the unstable case exhibits synchronized variations in instantaneous pressure and heat release over time. This in-phase variation establishes a feedback loop that sustains the instability. Figure 26(b) compares instantaneous pressure and heat release frequency at the CP1 probe location for unstable cases. Both exhibit a dominant frequency of 4.3 kHz, indicating a feedback mechanism in the unstable case. The heat release frequency aligns with the first longitudinal mode of the combustor. This positive feedback loop between heat release and acoustic modes amplifies pressure fluctuations in low-temperature cases. The trigger for instability is further analyzed by examining the potential delay in combustion caused by low-temperature methane injection. Figure 27 illustrates the variation of pressure and heat release at the CP1 probe for unstable and stable simulations over a shorter time window. In the unstable case (Figure 27(a)), a noticeable delay of 0.06ms is observed between the pressure rise and the heat release peak, as highlighted in the figure. This delay is absent in the stable case, as shown in Figure 27(b). The source of the delay in the unstable case is linked to the onset of longitudinal wave instability in the combustor.

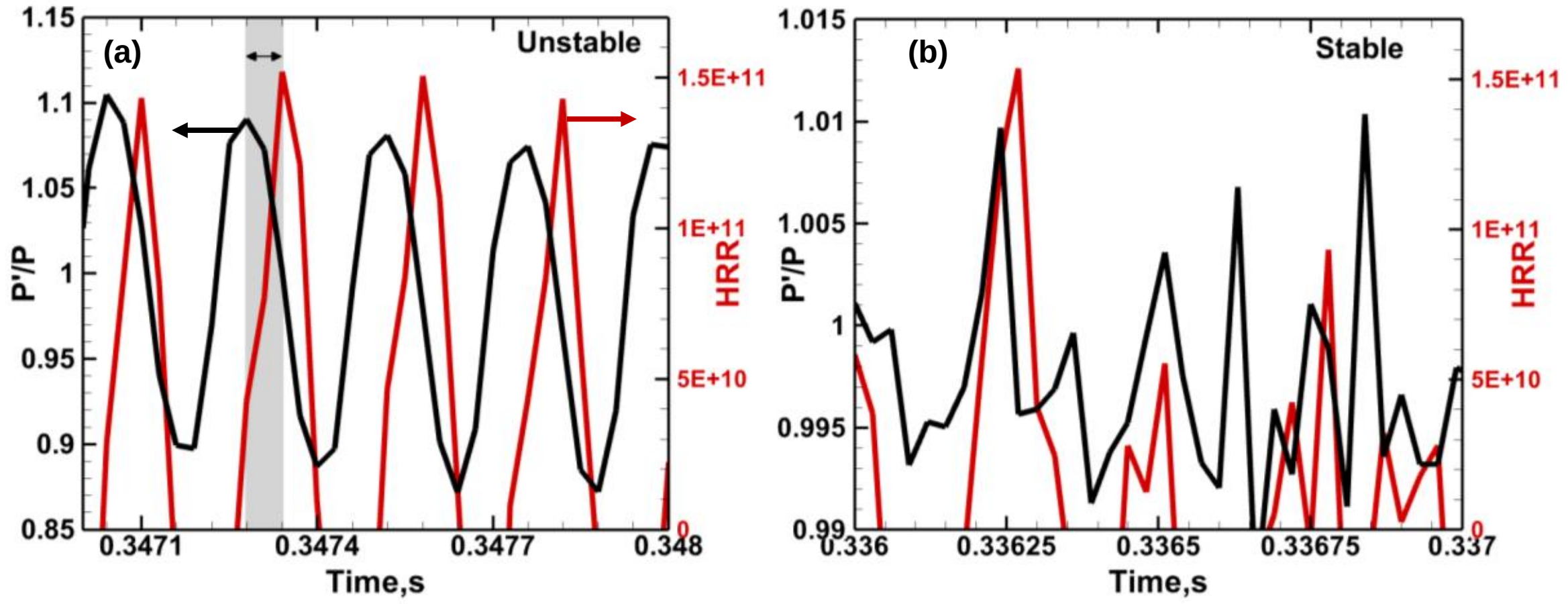

**FIG.27.** P'-Q' variation for (a) unstable and (b) stable cases

Figure 28 illustrates the pressure evolution alongside axial velocity at the CP1 probe location. The pressure-axial velocity variation indicates that during a high-pressure phase at CP1, the axial velocity drops below zero, signifying upstream flow movement during pressure wave arrival at the combustor head end. Figure 28(b) highlights the reversal in axial velocity, which peaks during the formation of a low-pressure zone at the combustor start. This periodic flow interruption disrupts the inflow into the combustor. The delays caused by these flow dynamics contribute to the heat release delay, as observed in Figure 27(a).

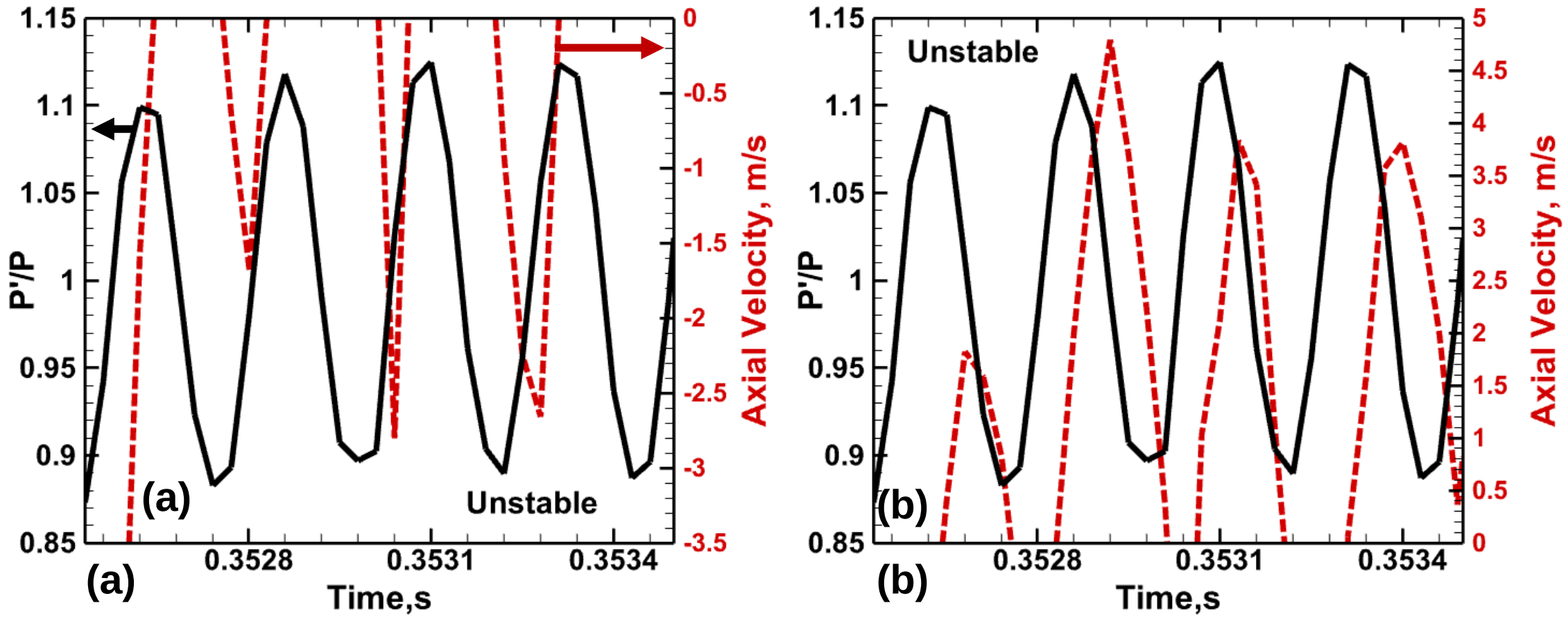

**FIG.28.** Pressure and axial velocity variation at CP1 location: (a) during wave arrival, and (b) wave departure

The instability mechanism is further analyzed by examining physical properties like specific heat and sound speed. Figure 29 shows specific heat variations at the exit of the methane pathway for both unstable and stable cases. In the unstable case (a), specific heat displays a nonlinear variation, while the stable case (b) maintains a near-constant value. Thermodynamic analysis reveals a nonlinear rise in specific heat as methane approaches its critical temperature (190K). As observed in the unstable case, rapid fluctuations in specific heat can induce high thermal gradients in the shear layer, leading to hydrodynamic disturbances. These disturbances can initiate and reinforce acoustic instability within the combustor.

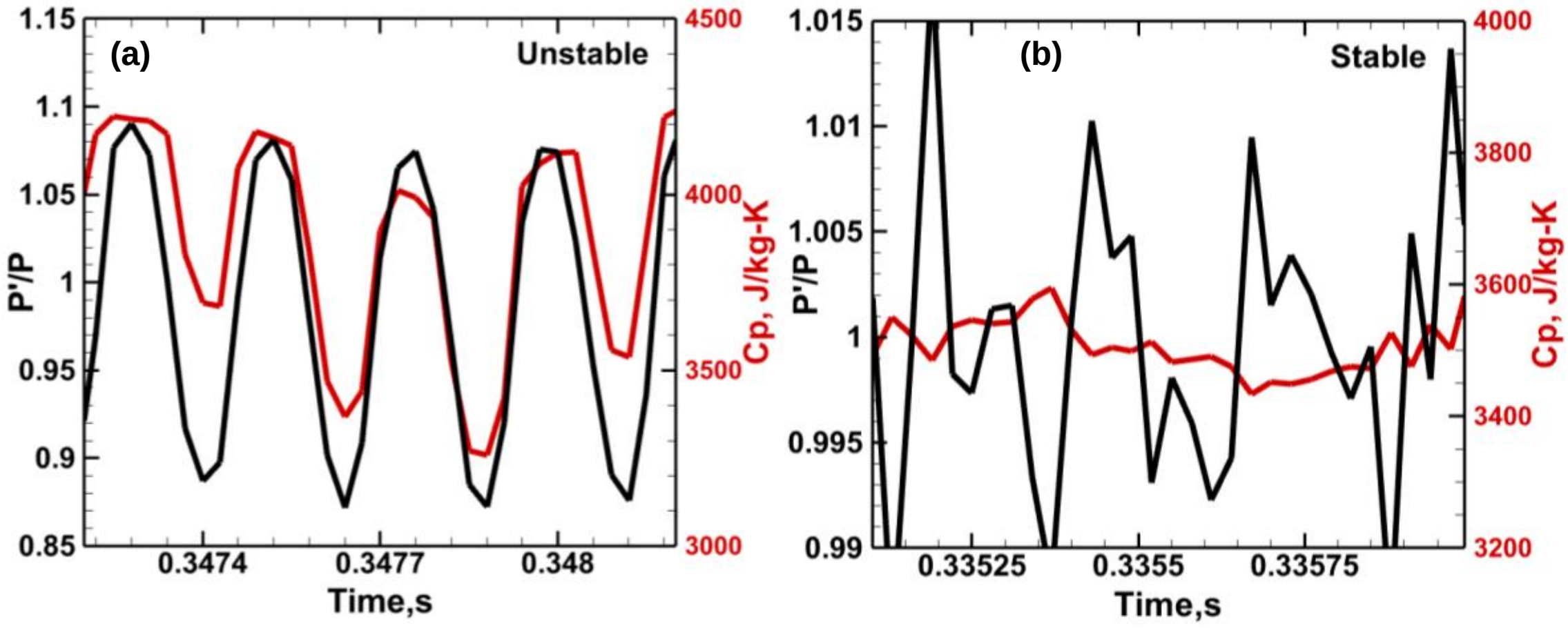

**FIG.29.** Specific heat variation in (a) unstable and (b) stable case

We illustrate that pressure-heat release coupling, periodic fuel flow interruptions, and nonlinear changes in thermodynamic properties can trigger and sustain instability. Low-temperature fuel slows chemical reactions, reducing fuel consumption rates and causing unburned gas to accumulate downstream of the injector faceplate. This accumulation and rapid heat release increase the potential for pressure wave development. Sudden heat release from unburned gas near the injector can trigger these pressure waves. Observations indicate that significant heat release occurring at the pressure anti-node of the longitudinal acoustic mode near the injector exit during every cycle of the acoustic oscillation can contribute to sustaining this instability.

## G. Shear Layer Dynamics

Shear layer dynamics are critical in understanding acoustic instability in rocket combustion chambers. Shearing distinct fuel and oxidizer streams from discrete injectors can lead to hydrodynamic instability. The ratio of fuel to oxidizer injection velocities significantly affects the characteristics of the shear layer, as even small perturbations within this layer can escalate into large-scale instabilities. These instabilities can amplify when the shear layer resonates with the acoustic modes of the combustion chamber. Pfitzner et al.[51] identified the density gradient as a critical parameter influencing shear layer dynamics. A lower velocity ratio can trigger changes in shear layer behavior by reducing the intensity of vortices and turbulence, resulting in slower and less efficient mixing. This scenario increases the potential for localized heat release fluctuations, contributing to combustion instability. In this section, we illustrate instability dynamics based on observations of the ratio of fuel to oxidizer injection velocities, commonly known as the velocity ratio, in rocket propulsion. This ratio is considered a vital stability parameter. The influence of the velocity ratio is examined across different temperature cases simulated. In particular, the role of the velocity ratio (fuel velocity, $V_f$ /oxidizer velocity, $V_o$) in triggering instability in low-temperature cases is identified. Bijukumar[52] found that a low hydrogen injection velocity plays a key role in triggering instability in LOx-$H_2$ engines. The study reveals that hydrogen injection velocity, rather than temperature, is a critical parameter for instability.

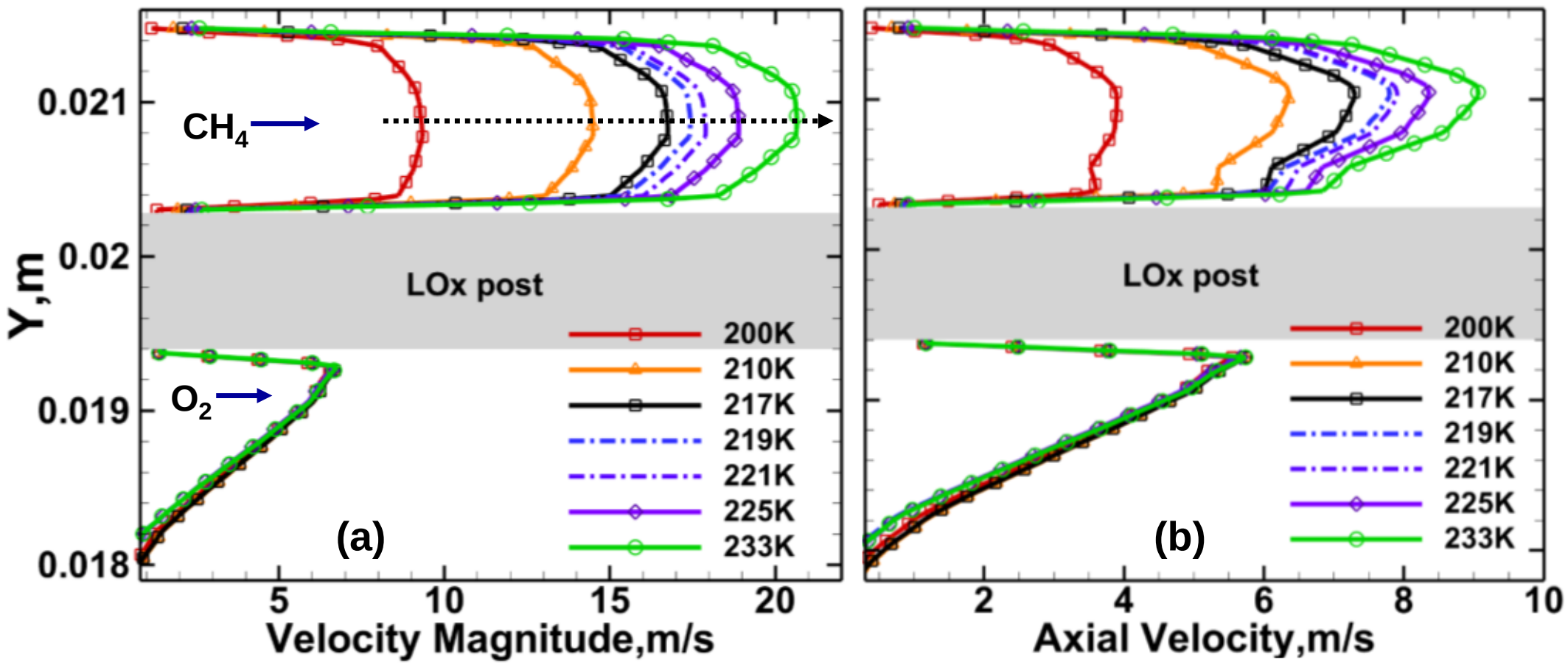

**FIG.30.** Variation in shear layer for different temperature cases (a) Velocity magnitude, and (b) axial velocity

Figure 30 illustrates the velocity distribution in the shear layer for different temperature cases. An increasing trend in methane velocity is observed with rising injection temperatures. A direct effect of injection temperature on velocity is evident, with a minimum velocity at an injection temperature of 200 K and a maximum peak velocity at 233 K. The axial velocity variation, as shown in Figure 30(b), follows a similar pattern. Since the simulations are conducted at the same mass flow rate, changes in fuel injection temperature directly affect the fuel injection velocity. As discussed, a transition to an unstable state occurs when the fuel temperature drops below 220 K. The dotted lines in the plots indicate the fuel velocity condition at which dynamic instabilities erupt in the combustor. This transition is explained in terms of the velocity ratio, as variations in fuel velocity alter the velocity ratio, thereby influencing the shear layer dynamics. A stability map based on these fuel temperature simulations is generated for the current combustor configuration, which has potential applicability to similar systems. Figure 31(a) presents a bar chart showing the peak-to-peak pressure fluctuation amplitude in the combustor for different fuel injection temperatures. The results indicate more excellent stability (lower pressure amplitudes) at higher fuel injection temperatures, which correspond to higher velocity ratios. Figure 31(b) illustrates the variation of instability with velocity ratio, showing a sharp decrease in instability amplitude as the velocity ratio increases. A critical velocity ratio of 3.8 is identified, below which the combustor transitions to an unstable state. This analysis demonstrates that a lower velocity ratio adversely affects shear layer dynamics, resulting in poor mixing, fuel accumulation, and increased boundary layer effects in the combustor's back-step region. These effects lead to uneven heat release and contribute to combustion instability at lower fuel injection temperatures.

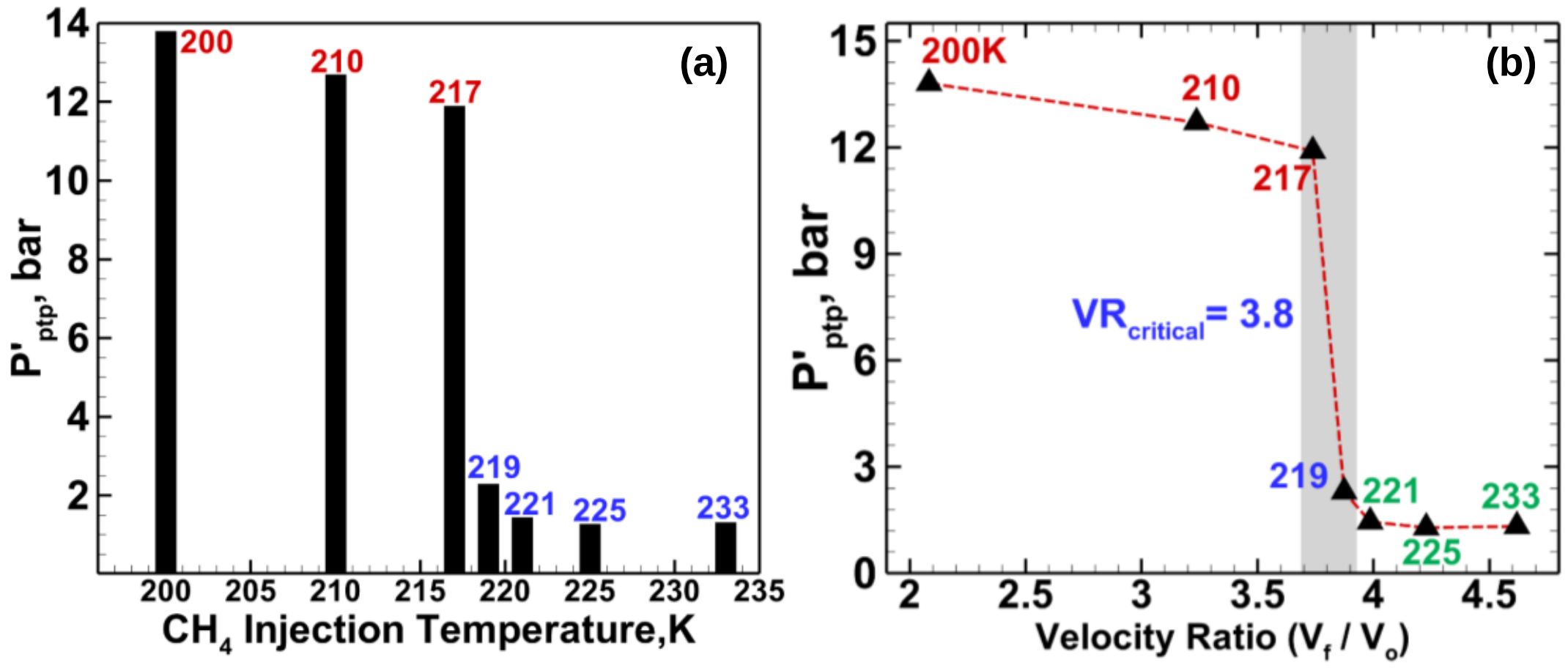

**FIG.31.** Stability map with (a) injection temperature and (b) velocity ratio

To confirm the impact of the velocity ratio on the stability of the combustor, we simulated two additional cases: one with a lower velocity ratio at a higher fuel injection temperature of 233 K, and the other with a higher velocity ratio at a lower fuel injection temperature of 200 K. The combustor's overall oxidizer-to-fuel (O/F) ratio has been adjusted while maintaining the same chamber pressure as in the earlier simulations to change the velocity ratio for these cases. Figure 32 displays the axial velocity plot for these additional cases: 200 K at an O/F ratio of 1.48 (denoted as 200KMR1.48) and 233K at an O/F ratio of 6.6 (denoted as 233KMR6.6). The axial velocity is plotted alongside the earlier unstable 200 K and stable 233 K cases. We observed the impact of a lower velocity ratio in the 233 K case, which exhibited unstable operation due to the lower fuel velocity than the oxidizer velocity, as shown in Figure 31. However, this instability could result from the lower velocity ratio and the higher O/F ratio in the combustor. In the 200KMR1.48 case, the velocity ratio increased considerably due to the higher methane flow rate, but the simulation still displayed unstable operation, similar to the earlier 200 K nominal O/F case. The higher fuel velocity (higher velocity ratio) was insufficient to suppress the instability associated with the low fuel injection temperature.

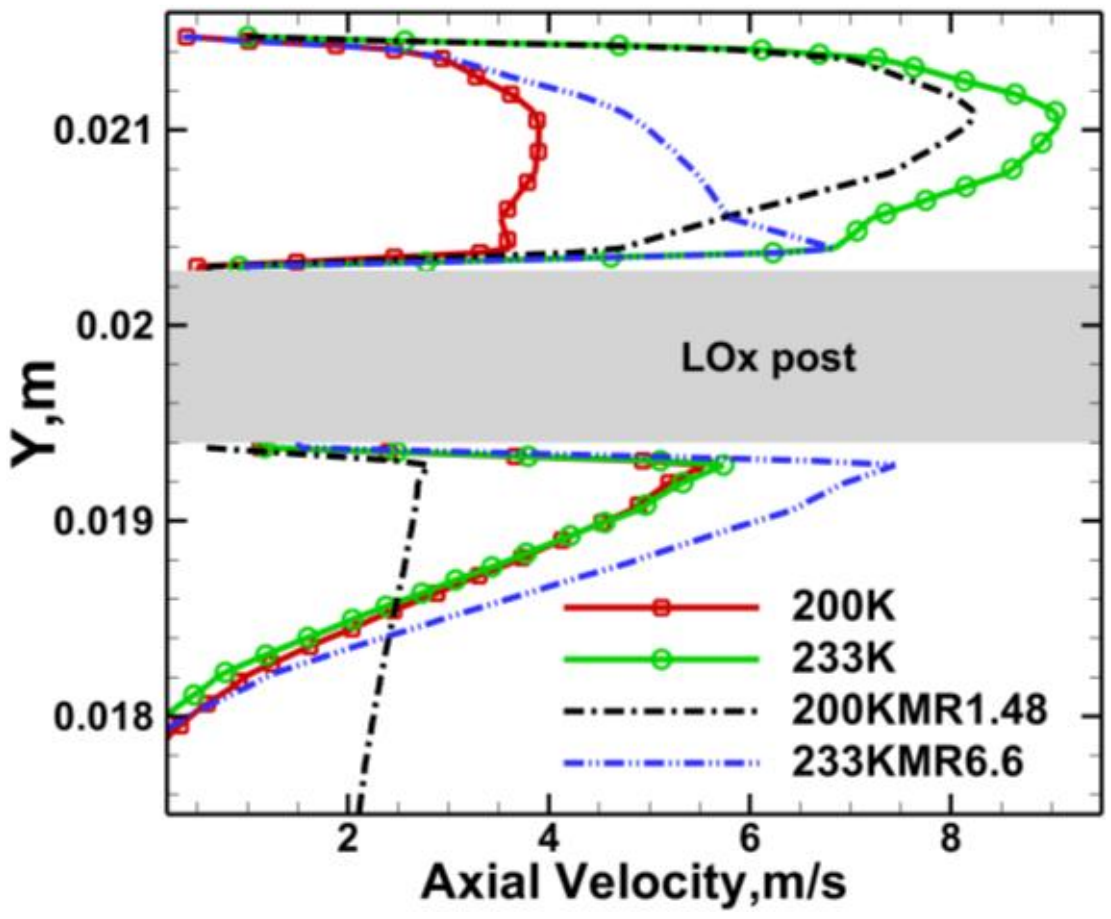

**FIG.32.** Axial velocity variation in shear layer for additional mixture ratio cases

## H. LOx post recess

The impact of a LOx post recess is also explored to mitigate the instability observed at low-temperature conditions. Figure 33 depicts the nominal LOx path and the recessed LOx path simulated. A 1D recess is introduced in this LES, where D denotes the LOx path diameter. Generally, recess is considered to improve combustion performance and stability. However, counter-observations have been reported in the work on the BKD chamber[53], where a recess adversely affected combustion stability. Specifically, it was observed that injectors without a LOx post recess were stable, while instability occurred with the recessed injector. This contradicts earlier studies on LOx-hydrogen tests.

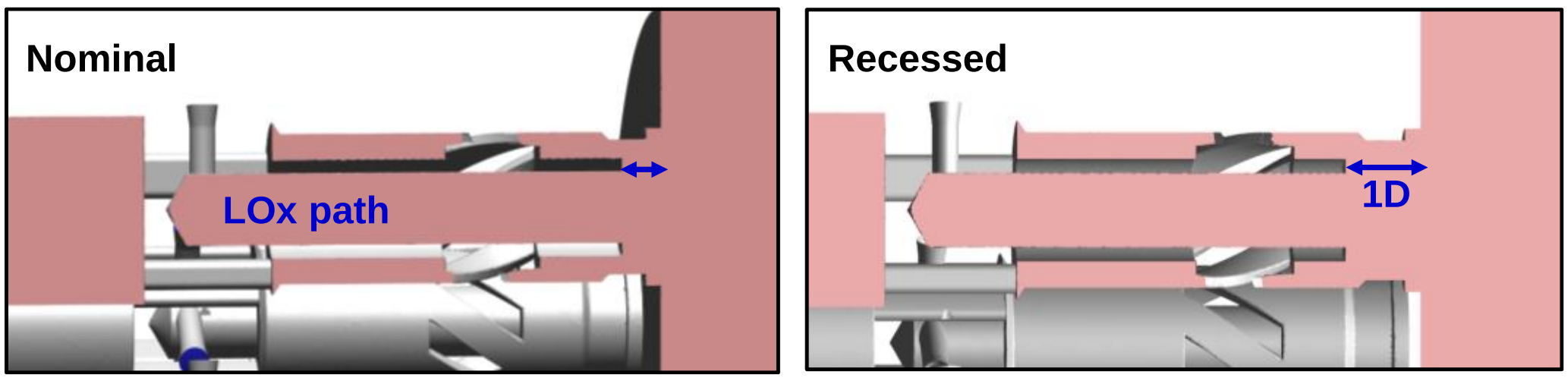

**FIG.33.** LOx post for nominal and recessed LES We conducted an LES simulation for the lowest methane injection temperature of 200 K to examine the impact of the recess on the high-amplitude pressure oscillations observed in the nominal recess case. The LES with recess at 200 K also showed the triggering of instability and the evolution to limit cycle oscillations, with a peak-to-peak amplitude of 15 bar. The introduction of the recess did not alter the instability scenario, which may be attributed to the extreme case of 200 K methane injection temperature. Further analysis of the impact of the recess on stability falls outside the scope of this work and will be addressed in future studies.

## VI. CONCLUSION

The paper investigates the supercritical combustion dynamics of a multi-element LOx-methane combustor operating under rocket-relevant conditions. A validated high-fidelity numerical framework is developed, incorporating real-gas thermodynamics, high-pressure transport physics, and flamelet-generated manifold combustion closure. A comprehensive study is conducted to understand the impact of fuel injection temperature on combustion stability. The study captures the intense acoustic activity in the combustor at lower fuel injection temperatures. It reveals highly unstable flame behavior under these conditions, contrasting with the stable characteristics observed at nominal fuel injection temperatures. Our analysis identifies unstable longitudinal acoustic modes (1L and its harmonics) in the combustor and successfully reproduces self-sustained limit cycle oscillations. A stability boundary temperature of 219 K is determined, below which the combustor becomes unstable. A detailed spectral and dynamic mode decomposition (DMD) analysis of the unstable case highlights the onset of longitudinal acoustic waves in the combustor. A comparative analysis between stable and unstable operations reveals fundamental differences in their behavior. The instability mechanism, illustrated through shear layer dynamics, emphasizes the critical role of the velocity ratio in triggering instability at low-temperature conditions. The low-temperature cases exhibit significantly lower velocity ratios compared to nominal temperature operations. A critical velocity ratio is identified below, which instability can trigger. A stability boundary temperature is identified for this combustor configuration; however, the limiting temperature can also apply to other LOx-methane engine designs. These findings are instrumental in optimizing high-pressure LOx-methane engine design and provide clear insights for its stable operation. Our work demonstrates the utility of a high-fidelity LES framework with decomposition tools like DMD to provide valuable insights into the impact of fuel injection temperature on combustion stability. It is the first-of-its-kind study to simulate the onset of combustion instability in a LOx-methane rocket-scale combustor operating at a lower methane injection temperature. The developed framework will be further tested to evaluate the combustor's stability under perturbed or off-nominal flow conditions. Future efforts will focus on determining the injector geometric features required for stable operation at lower fuel injection temperatures.

## ACKNOWLEDGEMENTS

The authors acknowledge the availability of 100 TeraFlop High-Performance Computing Facility at the Liquid Propulsion Systems Center (LPSC), featuring direct water-cooled processor technology. We thank M/s StarOneIT, Trivandrum, for their logistics support in maintaining this facility.

## AUTHOR DECLARATIONS

### CONFLICT OF INTEREST

The authors have no conflicts to disclose.

### DATA AVAILABILITY

The data that support the findings of this study are available from the corresponding author upon reasonable request.

## APPENDIX A:

## Validation studies:

The first validation test of the computational framework is carried out on a well-known Mascotte chamber experimented by Candel and Singla et al.[54]. The facility was originally developed for hydrogen-oxygen combustion but was later extended to enable advanced experimental investigations on LOx-methane combustion. It serves as a relevant benchmark case for testing the numerical framework under rocket-relevant operating conditions. The Mascotte chamber has a rectangular cross-section of a 50 × 50 $mm^2$ square duct and a length of 400mm, which ends with a converging-diverging nozzle. To reduce the computational expense of LES, we invoked a truncated domain of 200mm in this study. Figure 34 (a) displays a rectangular computational domain with a shear coaxial injector and representative flame interface at the entry. The geometric details of the injector and chamber are similar, as reported by Kim et al.[55] . The injector dimensions d1, d2, d3, and d4 parameters can be referred from our earlier work[26]. Figure 34 (a) shows a zoomed view of the injector with geometric features like LOx post thickness and post divergence appropriately modeled without simplification. Figure 35 compares the OH concentration between the LES and G2 test cases. In the G2 test, emission imaging of the excited OH radical was recorded and time-averaged to determine the mean flame structure. Only a qualitative picture of OH concentration is available from the test[56]. The bottom part of Figure 34(b) displays the Abel-transformed OH concentration image in the G2 test, whereas the sampled OH concentration from LES is shown at the top.

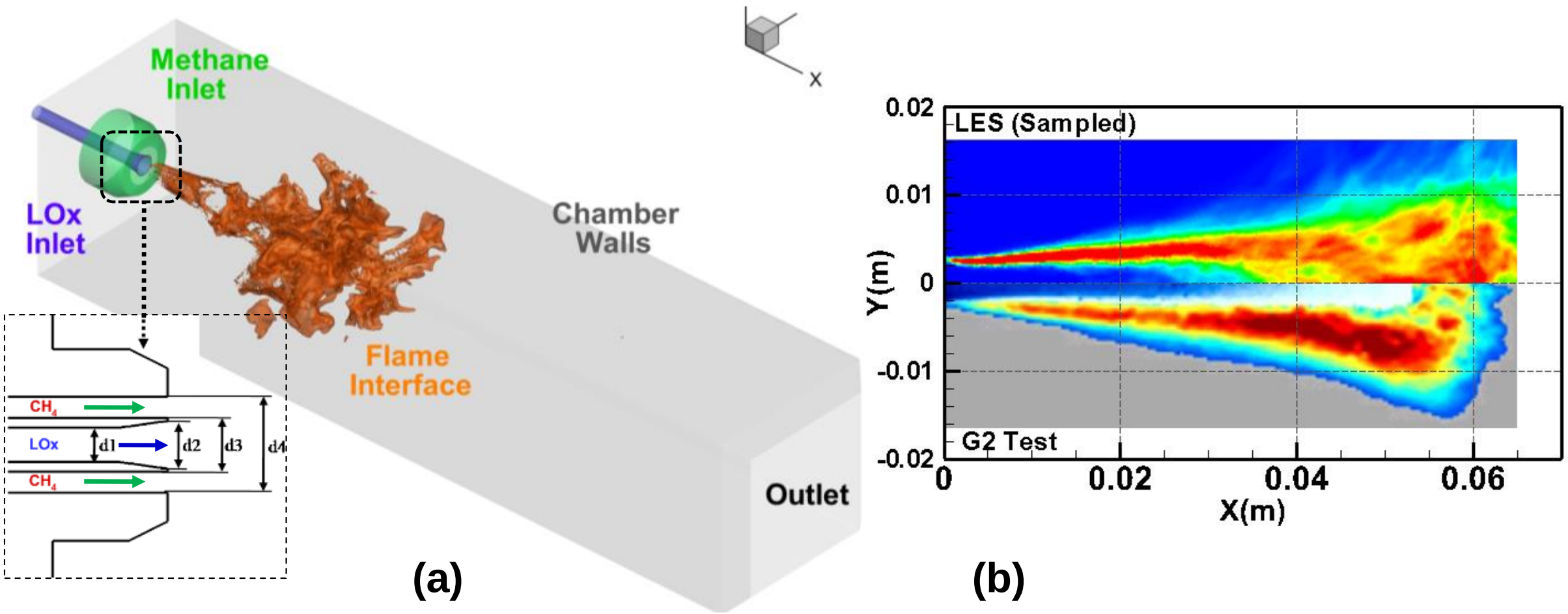

**FIG. 34.** (a)Computational domain of G2 test chamber (representative flame) with a zoomed view of the injector & (b) OH concentration comparison with G2 test case

A good agreement in flame length and radial expansion is observed between the LES and the test results. In both cases, the flame terminates at an axial distance of 60 mm from the injector exit. While some profile differences are noticed towards the end of the flame, the LES effectively captures key phenomena, including flame anchoring, shear-layer combustion, radial expansion driven by transcritical oxygen pseudo-boiling, and the overall flame potential core. This close alignment with the benchmark test satisfies the primary goal of the first validation study.

The second validation test was carried out on a self-excited combustor of Purdue University developed by Yu et al.[14] In this validation study, thermo-acoustic simulation of the continuously variable resonance combustor (CVRC) is conducted for an oxidizer post length of 12 cm using the large eddy simulation framework as reported in our earlier work[33]. Figure 35 illustrates the geometric details of the combustor, including the oxidizer slotted plunger, fuel inlet manifold, and injector geometry. A close-up view of the oxidizer slotted plunger highlights the 36 orifices through which fuel enters the domain.

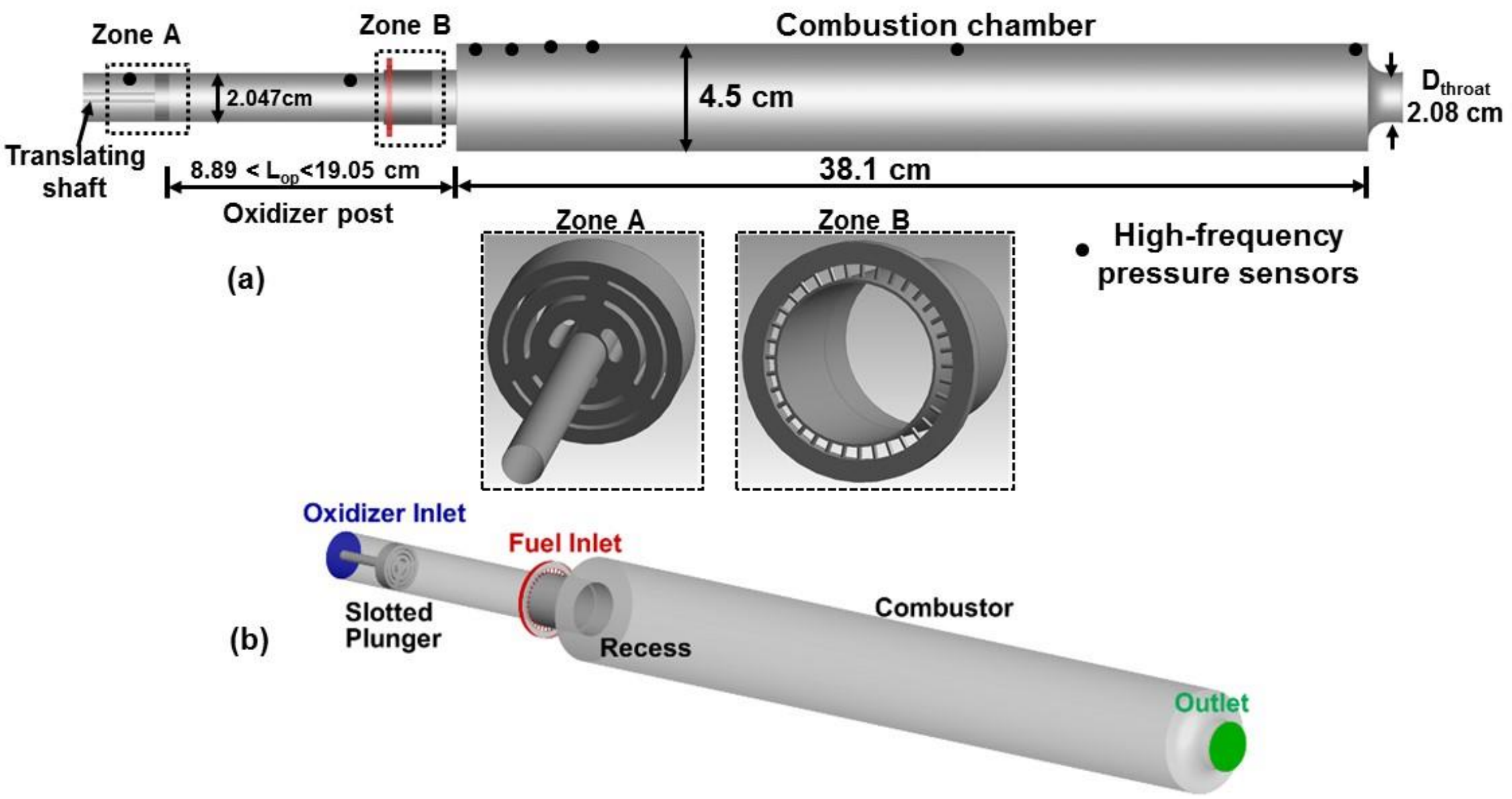

**FIG. 35.** Close up view of oxidizer plunger (a) & fuel Inlet Section (b) Computational domain CVRC

Figure 36 (a) displays the comparison of the test to simulation pressure variation at the combustor endpoint. The LES pressure variation displays the self-excited limit cycle behavior, which closely matches the test data. A peak-to-peak pressure fluctuation amplitude of 5 bar in the simulation was noticed, which amounts to 30% of the mean chamber pressure. A reasonable match is observed between the test and simulation, with an underprediction of pressure amplitude on the positive side of the spectrum. Figure 36(b) displays a spectral plot of pressure data, which depicts the harmonics of the first longitudinal mode in the combustor. The spectral plot shows a well-organized pressure oscillation in the combustor. LES displays a 1L mode at 1520Hz, with second harmonics (2L) present near 3100Hz and so on, while the test exhibits a 1L mode close to 1400Hz [14]

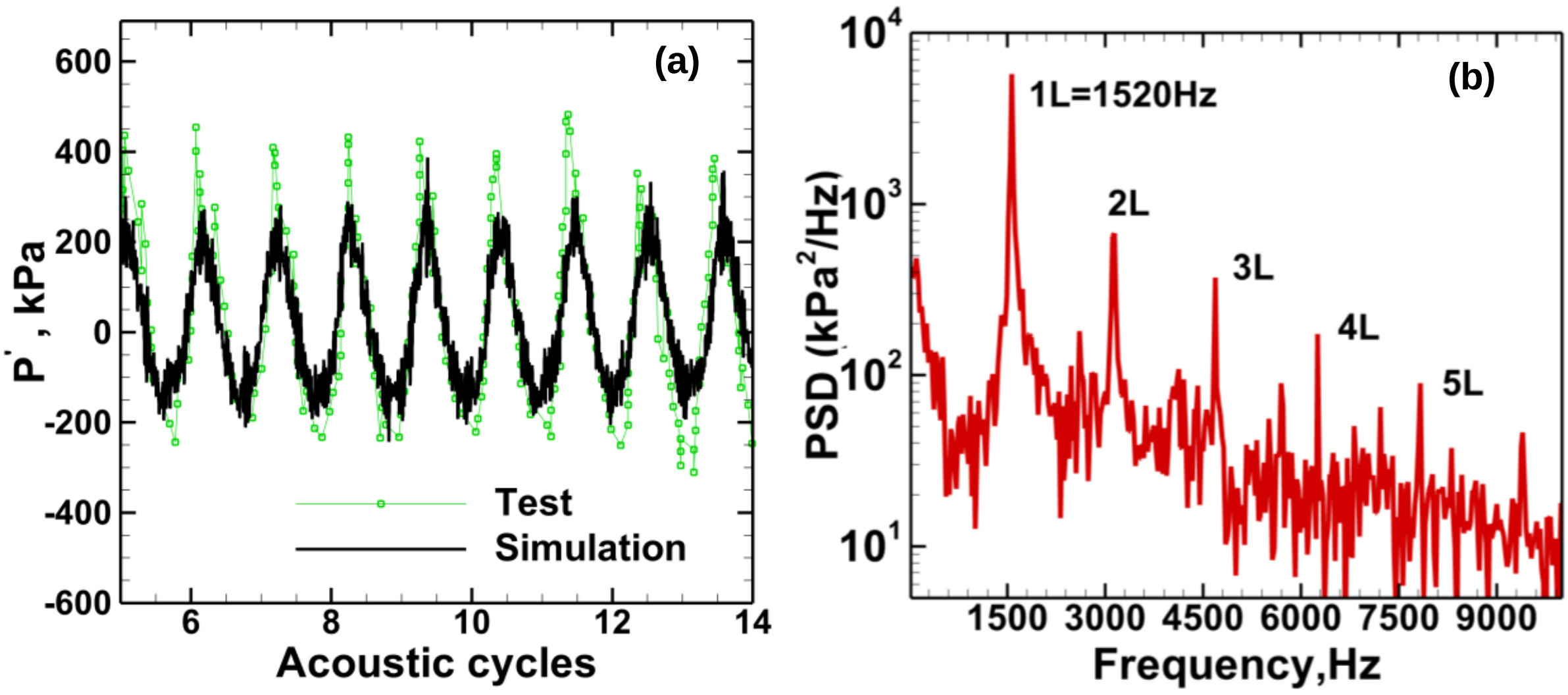

**FIG. 36.** (a) Peak-to-peak raw pressure fluctuations. Test results are taken from[57]
(b) Harmonics of raw pressure data

This validation study demonstrates satisfactory agreement in pressure fluctuation amplitude and dominant frequency between LES and test data. It highlights the effective application of the LES methodology in capturing self-sustained instabilities in CVRC. This validated methodology is subsequently utilized to investigate the impact of lower fuel injection temperatures on the stability of LOx-methane supercritical flames.

## REFERENCES

[1] Rayleigh, "The explanation of certain acoustical phenomena," Nature **18**(455), 319–321 (1878).

[2] J. Hulka, and J.J. Hutt, "Instability phenomena in liquid oxygen/hydrogen propellant rocket engines," PROGRESS IN ASTRONAUTICS AND AERONAUTICS **169**, 39–72 (1995).

[3] Y. Nunome, T. Onodera, M. Sasaki, T. Tomita, K. Kobayashi, and Y. Daimon, "Combustion Instability Phenomena Observed During Cryogenic Hydrogen Injection Temperature Ramping Tests for Single Coaxial Injector Elements," in *47th AIAA/ASME/SAE/ASEE Joint Propulsion Conference & Exhibit*, (American Institute of Aeronautics and Astronautics, Reston, Virigina, 2011), p. 6027.

[4] J.P. Wanhainen, H.C. Parish, and E.W. Conrad, *Effect of Propellant Injection Velocity on Screech in 20,000-Pound Hydrogen-Oxygen Rocket Engine* (National Aeronautics and Space Administration, 1966).

[5] S. Gröning, J. Hardi, D. Suslov, and M. Oschwald, "Influence of hydrogen temperature on the stability of a rocket engine combustor operated with hydrogen and oxygen: A new hydrogen temperature ramping experiment," CEAS Space Journal **9**(1), 59–76 (2017).

[6] R.J. Jensen, H.C. Dodson, and S.E. Claflin, *LOX/Hydrocarbon Combustion Instability Investigation* (1989).

[7] J. Martin, W. Armbruster, R. Stützer, D. Suslov, J. Hardi, and M. Oschwald, "Flame dynamics of an injection element operated with LOX/H2, LOX/CNG and LOX/LNG in a sub- and supercritical rocket combustor with large optical access," International Journal of Spray and Combustion Dynamics **15**(3), 147–165 (2023).

[8] H. Kawashima, K. KOBAYASHI, T. Tomita, and T. Kaneko, "A Combustion Instability Phenomenon on a LOX/Methane Subscale combustor," in *46th AIAA/ASME/SAE/ASEE Joint Propulsion Conference & Exhibit*, (American Institute of Aeronautics and Astronautics, Reston, Virigina, 2010).

[9] N. Zhang, W. WANG, and J. SUN, "Demonstration of a 600kN class lox/methane rocket engine," in *63th International Aeronautical Congress*, (2012).

[10] L. CROCCO, and C.E. MITCHELL, "Nonlinear Periodic Oscillations in Rocket Motors with Distributed Combustion," Combustion Science and Technology **1**(2), 146–169 (1969).

[11] C. L., "Aspects of Combustion Instability in Liquid Propellant Rocket Motors," J. Amer. Rocket Soc. **21**, 163 (1951).

[12] D.T. Harrje, and F.H. Reardon, *Liquid Propellant Rocket Combustion Instability. NASA SP-194* (1972).

[13] F.E.C. Culick, "Combustion Instabilities in Propulsion Systems," in *Unsteady Combustion*, edited by M.V. and W.J.H. Culick F. and Heitor, (Springer Netherlands, Dordrecht, 1996), pp. 173–241.

[14] Y. Yu, J.C. Sisco, S. Rosen, A. Madhav, and W.E. Anderson, "Spontaneous Longitudinal Combustion Instability in a Continuously-Variable Resonance Combustor," J Propuls Power **28**(5), 876–887 (2012).

[15] S. Gröning, J. Hardi, D. Suslov, and M. Oschwald, "Injector-Driven Combustion Instabilities in a Hydrogen/Oxygen Rocket Combustor," J Propuls Power **32**, 1–14 (2016).

[16] W. Armbruster, J.S. Hardi, D. Suslov, and M. Oschwald, "Experimental investigation of self-excited combustion instabilities with injection coupling in a cryogenic rocket combustor," Acta Astronaut **151**, 655–667 (2018).

[17] M.E. Harvazinski, C. Huang, V. Sankaran, T.W. Feldman, W.E. Anderson, C.L. Merkle, and D.G. Talley, "Coupling between hydrodynamics, acoustics, and heat release in a self-excited unstable combustor," Physics of Fluids **27**(4), 045102 (2015).

[18] A. Urbano, L. Selle, G. Staffelbach, B. Cuenot, T. Schmitt, S. Ducruix, and S. Candel, "Exploration of combustion instability triggering using Large Eddy Simulation of a multiple injector liquid rocket engine," Combust Flame **169**, 129–140 (2016).

[19] R. Garby, L. Selle, and T. Poinsot, "Large-Eddy Simulation of combustion instabilities in a variable-length combustor," Comptes Rendus Mécanique **341**(1), 220–229 (2013).

[20] A. Sharma, A. De, and S.S. Kumar, "Numerical investigation of supercritical combustion dynamics in a multi-element LOx–methane combustor using flamelet-generated manifold approach," Physics of Fluids **35**(11), 115125 (2023).

[21] T. Schmitt, G. Staffelbach, S. Ducruix, S. Gröning, J. Hardi, and M. Oschwald, "Large-Eddy Simulations of a sub-scale liquid rocket combustor: influence of fuel injection temperature on thermo-acoustic stability," (2017).

[22] T. Schmitt, A. Coussement, S. Ducruix, and S. Candel, "Large Eddy Simulations of high amplitude self-sustained acoustic oscillations in a rocket engine coaxial injector in the transcritical regime," in *Space Propulsion, Roma, Italy*, (2016).

[23] T. Ota, H. Terashima, and N. Oshima, "Mechanisms for severe combustion instabilities induced by low-temperature fuel injection of an H2/O2 rocket-type combustor," in *AIAA SCITECH 2022 Forum*, (American Institute of Aeronautics and Astronautics, Reston, Virginia, 2022).

[24] M. Harvazinski, and T. Shimizu, *Computational Investigation on the Effect of the Oxidizer Inlet Temperature on Combustion Instability* (2019).

[25] J.C. Oefelein, and V. Yang, "Modeling High-Pressure Mixing and Combustion Processes in Liquid Rocket Engines," J Propuls Power **14**(5), 843–857 (1998).

[26] A. Sharma, A. De, and S. Kumar, "Assessment of Statistical and Kinetics based models for LOx-Methane Green Propellant Combustion," International Journal of Energy for a Clean Environment **24**(7), 1–57 (2022).

[27] G. Singla, P. Scouflaire, C. Rolon, and S. Candel, "Transcritical oxygen/transcritical or supercritical methane combustion," Proceedings of the Combustion Institute **30**(2), 2921–2928 (2005).

[28] S. Candel, M. Juniper, G. Singla, P. Scouflaire, and C. Rolon, "Structure and dynamics of cryogenic flames at supercritical pressure," Combustion Science and Technology **178**(1–3), 161–192 (2006).

[29] S. Boulal, N. Fdida, L. Matuszewski, L. Vingert, and M. Martin-Benito, "Flame dynamics of a subscale rocket combustor operating with gaseous methane and gaseous, subcritical or transcritical oxygen," Combust Flame **242**, 112179 (2022).

[30] A.S.A.V.T.J.T. Rahul Kumar, "Thermodynamic Modelling of High-Pressure LOx-Methane Rocket Engine," in *FMFP2023, IITJodhpur, India*, (Fluid Mechanics and Fluid Power, Jodhpur, 2024).

[31] B.E. Poling, J.M. Prausnitz, and J.P. O'Connell, *Properties of Gases and Liquids*, Fifth edition. (McGraw-Hill Education, New York, 2001).

[32] T.H. Chung, M. Ajlan, L.L. Lee, and K.E. Starling, "Generalized multiparameter correlation for nonpolar and polar fluid transport properties," Ind Eng Chem Res **27**(4), 671–679 (1988).

[33] A. Sharma, A. De, and S. Kumar, "Investigation of injector-coupled combustion dynamics in a methane–oxygen combustor using large eddy simulation and dynamic mode decomposition," Physics of Fluids **36**(6), (2024).

[34] A. Sharma, A. De, and S.S. Kumar, "Analysis of Self-Excited Combustion Instability in a Sub-scale Rocket Engine Using Large Eddy Simulation," in *Proceeding of Proceedings of the 26thNational and 4th International ISHMT-ASTFE Heat and Mass Transfer Conference December 17-20, 2021, IIT Madras, Chennai-600036, Tamil Nadu, India*, (Begellhouse, Connecticut, 2022), pp. 637–643.

[35] A. Sharma, A. De, V.M. Thannickal, T. John Tharakan, and S. Sunil Kumar, "Numerical Investigation of Combustion Dynamics in a Multi-element Combustor Using Flamelet Approach," (2024), pp. 225–240.

[36]"ANSYS Inc., 2021. 'Ansys Fluent User's Guide. Release 2021R1'. Canonsburg, USA.," (n.d.).

[37] J.A. van Oijen, and L.P.H. de Goey, "Modelling of premixed counterflow flames using the flamelet-generated manifold method," Combustion Theory and Modelling **6**(3), 463–478 (2002).

[38] J.A. van Oijen, A. Donini, R.J.M. Bastiaans, J.H.M. ten Thije Boonkkamp, and L.P.H. de Goey, "State-of-the-art in premixed combustion modeling using flamelet generated manifolds," Prog Energy Combust Sci **57**, 30–74 (2016).

[39] R. Yadav, I. Verma, S. Orsino, P. Sharkey, P. Nakod, and S. Li, "Bluff-Body Stabilized Flame Simulations using SBES in Combination with the Flamelet Generated Manifold Combustion Model," in *AIAA Scitech 2019 Forum*, (American Institute of Aeronautics and Astronautics, 2019).

[40] R. Yadav, and P. Nakod, “Numerical Computation of a Turbulent Lifted Flame Using Flamelet Generated Manifold With Different Progress Variable Definitions,” in *ASME 2015 Gas Turbine India Conference*, (American Society of Mechanical Engineers, 2015).

[41] S. Patwardhan, P. Nakod, S. Orsino, and C. Arguinzoni, “Numerical Investigation of Combustion Instabilities in a Single Element Lean Direct Inject (LDI) Combustor Using Flamelet Based Approaches,” J Eng Gas Turbine Power **142**, (2020).

[42] B. Bhatia, A. De, D. Roekaerts, and A.R. Masri, “Numerical analysis of dilute methanol spray flames in vitiated coflow using extended flamelet generated manifold model,” Physics of Fluids **34**(7), 075111 (2022).

[43] V. Zhukov, and A. Kong, “A Compact Reaction Mechanism of Methane Oxidation at High Pressures,” Progress in Reaction Kinetics and Mechanism **43**, 62–78 (2018).

[44]“ANSYS Fluent Theory Guide, ANSYS, Inc., 275 Technology Drive Canonsburg, PA 15317, November 2020,” (n.d.).

[45] P.J. SCHMID, “Dynamic mode decomposition of numerical and experimental data,” J Fluid Mech **656**, 5–28 (2010).

[46] C. Huang, W. Anderson, M. Harvazinski, and V. Sankaran, “Analysis of Self-Excited Combustion Instabilities Using Decomposition Techniques,” AIAA Journal **54**, 1–17 (2016).

[47] T. Sayadi, and P.J. Schmid, “Parallel data-driven decomposition algorithm for large-scale datasets: with application to transitional boundary layers,” Theor Comput Fluid Dyn **30**(5), 415–428 (2016).

[48] P. Das, and A. De, “Numerical investigation of flow structures around a cylindrical afterbody under supersonic condition,” Aerosp Sci Technol **47**, 195–209 (2015).

[49] R.K. Soni, and A. De, “Role of jet spacing and strut geometry on the formation of large scale structures and mixing characteristics,” Physics of Fluids **30**(5), 056103 (2018).

[50] R.K. Soni, and A. De, “Investigation of mixing characteristics in strut injectors using modal decomposition,” Physics of Fluids **30**(1), 016108 (2018).

[51] A. Doehring, T. Trummler, M. Pfitzner, and M. Klein, “Two-dimensional direct numerical simulation study of multicomponent mixing with phase transition in a transcritical shear layer,” Physics of Fluids **36**(6), (2024).

[52] K.S Bijukumar, Role Of Hydrogen Injection Temperature On The Combustion Instability Of Cryogenic Rocket Engine, 2014.

[53] J. Martin, W. Armbruster, J.S. Hardi, D. Suslov, and M. Oschwald, “Experimental Investigation of Self-Excited Combustion Instabilities in a LOX/LNG Rocket Combustor,” J Propuls Power **37**(6), 944–951 (2021).

[54] S. Candel, M. Juniper, G. Singla, P. Scouflaire, and C. Rolon, “Structure and dynamics of cryogenic flames at supercritical pressure,” Combustion Science and Technology **178**(1–3), 161–192 (2006).

[55] T. Kim, Y. Kim, and S.K. Kim, "Effects of pressure and inlet temperature on coaxial gaseous methane/liquid oxygen turbulent jet flame under transcritical conditions," Journal of Supercritical Fluids **81**, 164–174 (2013).

[56] G. Singla, P. Scouflaire, C. Rolon, and S. Candel, "Transcritical oxygen/transcritical or supercritical methane combustion," Proceedings of the Combustion Institute **30 II**(2), 2921–2928 (2005).

[57] S. Srinivasan, R. Ranjan, and S. Menon, "Flame Dynamics During Combustion Instability in a High-Pressure, Shear-Coaxial Injector Combustor," Flow Turbul Combust **94**(1), 237–262 (2015).